# EUCA: the End-User-Centered Explainable AI Framework

WEINA JIN, Simon Fraser University
JIANYU FAN, Simon Fraser University
DIANE GROMALA, Simon Fraser University
PHILIPPE PASQUIER, Simon Fraser University
GHASSAN HAMARNEH, Simon Fraser University

The ability to explain decisions to end-users is a necessity to deploy AI as critical decision support. Yet making AI explainable to non-technical end-users is a relatively ignored and challenging problem. To bridge the gap, we first identify twelve end-user-friendly explanatory forms that do not require technical knowledge to comprehend, including feature-, example-, and rule-based explanations. We then instantiate the explanatory forms as prototyping cards in four AI-assisted critical decision-making tasks, and conduct a user study to co-design low-fidelity prototypes with 32 layperson participants. The results confirm the relevance of using explanatory forms as building blocks of explanations, and identify their proprieties — pros, cons, applicable explanation goals, and design implications. The explanatory forms, their proprieties, and prototyping supports (including a suggested prototyping process, design templates and exemplars, and associated algorithms to actualize explanatory forms) constitute the End-User-Centered explainable AI framework EUCA, and is available at http://weinajin.github.io/end-user-xai. It serves as a practical prototyping toolkit for HCI/AI practitioners and researchers to understand user requirements and build end-user-centered explainable AI.

## 1 INTRODUCTION

*Problem statement.* Doctors, judges, drivers, bankers, and other decision-makers require explanations from artificial intelligence (AI) when they use AI for critical decision support. As AI becomes pervasive in high-stakes decision-making tasks, such as in supporting medical, military, legal, and financial judgments, making AI explainable to its users is crucial to identify potential errors and establish trust [40]. The growing research community of eXplainable AI (XAI) aims to address such problems and "open the black box of AI" [32]. XAI literature generally divides its users into two groups according to their level of technical knowledge in AI: Technical users and non-technical users [19, 43, 77, 78, 81, 90]. The primary focus of current XAI research, however, is on debugging, understanding, and improving AI models for technical users, such as data scientists, AI researchers, and developers, leaving the largest and most diverse group of XAI users largely ignored: the non-technical end-users [19, 72]. Non-technical end-users, or end-users for short, can be either laypersons — such as drivers overseeing autonomous driving vehicles — or domain experts — such as doctors using AI-assisted technology in diagnostic tasks [25, 45, 48], judges using AI to support reaching a guilt verdict [54], and bankers using AI to assist in approving loan applications.

*Challenges.* Compared to developing XAI for technical users that mainly needs to deal with technical challenges in XAI [33], developing XAI for end-users faces even greater challenges: **1**) **No technical knowledge**: unlike technical users, end-users typically do not possess technical knowledge in AI, machine learning, data science, or programming, making some explanation methods which presume users' prior knowledge in AI (such as gradients, activations, neurons, layers) unviable. **2) Diverse user roles, tasks, and explanation goals**: when developing XAI for technical users, users have a relatively unified need: they utilize explanations mainly to debug, gain insights on the AI model, and improve it accordingly [46]. In contrast, developing XAI for end-users must adapt to the variability in the end-users' roles, tasks, and explanation goals. For example, a doctor may demand distinct explanations from AI when using it as a diagnostic support system, whereas a human resources specialist resorts to explanations to support her hiring decisions (different end-users and tasks). Even if an XAI system is built for the same task, users' explanation goal may vary. For example, a house seller may leverage the explanation on AI predictions to boost her property value, whereas a realtor may need the explanation to dig into why AI's prediction diverges from her own judgment.

*Research gaps.* Given these challenges, there is an urgent need for end-user-centered XAI design guidance to support AI practitioners' and researchers' XAI design process on critical decision support tasks. Although in recent years, we have witnessed booming research on XAI in both human-computer interaction (HCI) and AI communities [11, 32, 81], research on end-user-centered XAI is still at its infancy. The AI community lacks and calls for such user-centered perspective [65, 72]. In the HCI community, the existing user-centric XAI design guidance [60, 64, 71, 89] utilized a traditional user-centered approach informed by users' requirements only, which may lead to technically-unachievable solutions limited by current AI capacity or training data availability [98]. There also lacks work to support practitioners' prototyping, participatory design, and UX/UI (user interaction/user interface) design process, which are the most desired support identified by prior user studies with XAI design practitioners [60, 96, 97].

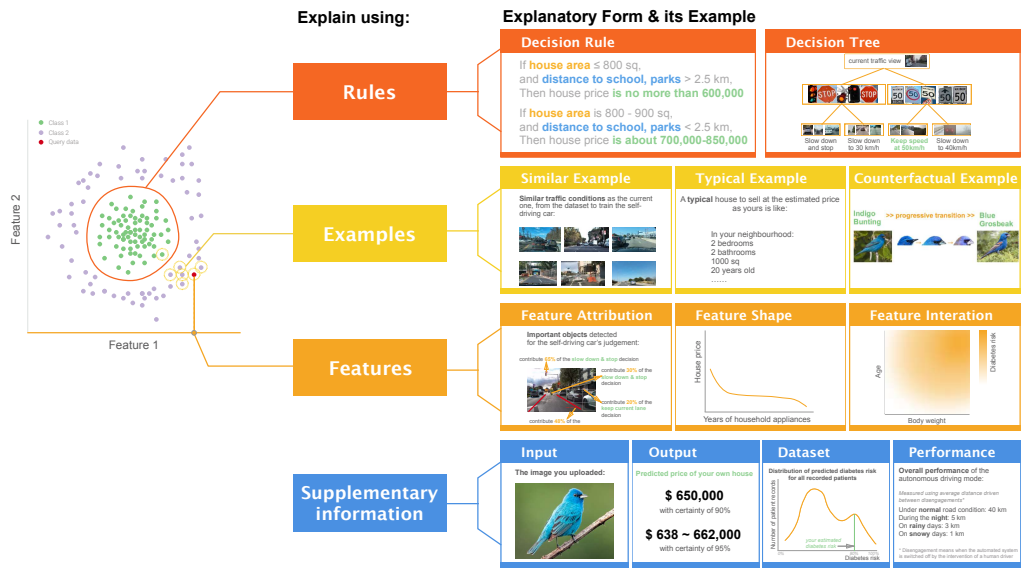

Fig. 1. **End-user-friendly explanatory forms in the EUCA framework**. EUCA consists of 12 end-user-friendly explanatory forms that are grouped into four categories: Explaining using rules, examples, features, and supplementary information. The former three categories correspond to the different aspects of showing AI's learned representations at the feature, instance, and decision boundary level, indicated on the left plot. Each form is instantiated by a prototyping card across four different tasks in our user study. The explanatory forms are a familiar language to both technical designers and end-users, thus overcoming the technical communication barriers between the two.

*Solution.* To address the above challenges and research gaps, we propose the End-User-Centered explainable AI framework EUCA. It is a practical prototyping framework to support the design and implementation process of end-user-centered XAI. EUCA applies to AI models for predictive tasks, such as classification and regression. The EUCA framework (Fig. 2) contains a suggested prototyping workflow (Fig. 3, Section 7.1.2), a set of end-user-friendly explanatory forms (Section 3) that consider both end-user literacy and technically-viable solutions, their design examples and templates (Fig. 1), their identified properties (pros, cons, applicable explanation goals, UI/UX design implications) from our user study (Section 5), and their associated XAI algorithms for implementation (Table 1). The full content of the framework is in the Appendix and on the EUCA website: http://weinajin.github.io/end-user-xai/). Table 1 summarizes its key messages.

The process of creating the EUCA framework is illustrated in Fig. 2. To tackle challenge **1**) **lack of technical knowledge**, we screened existing XAI techniques, summarized their final representation forms for explanation, and selected forms that do not require any prior technical knowledge to understand. We identified twelve end-user-friendly explanatory forms (Fig. 1): Explaining using **features** (including feature attribution, feature shape, and feature interaction), **examples** (similar, prototypical, and counterfactual example), **rules** (decision rule and decision tree), and some necessary **supplementary information** (input, output, dataset, performance). Since they are derived from technical works, the explanatory forms naturally link design representations to XAI algorithms. They enable designers to fully explore the technical feasible solution space, and not have to worry about whether their design solutions are technically feasible. Those forms are also a familiar language to the end-users, thus providing opportunities to involve users in the prototyping and participatory design process.

To address challenge **2**) **end-user's roles, tasks, and needs diversity**, EUCA incorporates user-centered design and suggests a prototyping process, so that users' context-specific requirements are fully understood and addressed in prototypes. To do so, we first instantiated the explanatory forms as prototyping cards on four AI-assisted critical decision-making tasks across health, safety, finance, and education, and conducted a user study with 32 layperson participants. The study consists of an interview and a card selection & sorting process, and it illustrates the idea of using explanatory forms as building blocks of explanation. Assisted by the prototyping cards, the participatory design process provides opportunities for designers and users to discuss and identify the optimal design together. Our user study also identified the strengths, weaknesses, applicable explanation goals, and UI/UX design implications for the twelve explanatory forms.

*Contribution.* The main contribution of EUCA is that it provides a practical prototyping framework for **AI practitioners** (UX designers, developers, etc.) to build end-user-centered XAI prototypes. The prototyping workflow and tangible design examples/templates support user-centered prototyping and co-design process, and enable end-users to communicate their context-specific explainability needs to practitioners. The suggested prototyping process (illustrated in Fig. 3) is intuitive to follow even for people outside the HCI/UX community. The explanatory forms are simple and familiar to both technical creators and non-technical users, allowing practitioners to invite all stakeholders to the co-design conversation. The coupled XAI algorithms support practitioners in implementing the low-fidelity prototypes to functional high-fidelity ones.

In addition to the above support for XAI practitioners, **HCI and AI researchers** may propose novel XAI interfaces/algorithms by using explanatory forms as building blocks and prototyping with EUCA. The user study findings uncover the strengths, weaknesses, and design implications for each explanatory form, providing opportunities to improve and create new explanation methods/interfaces.

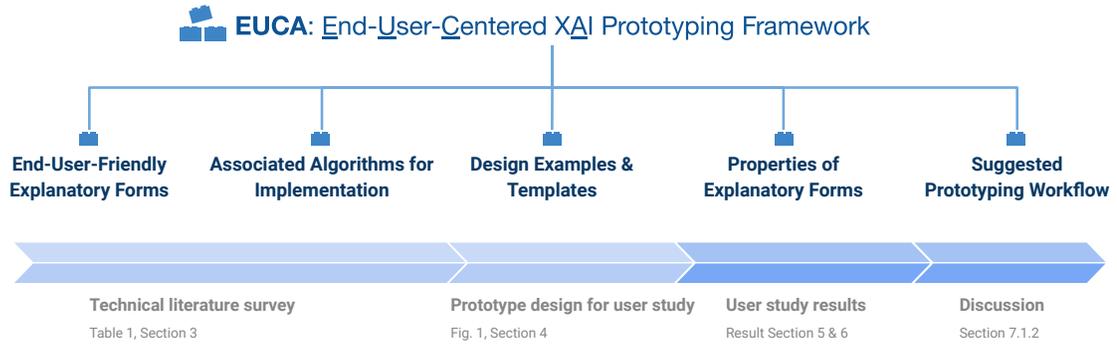

Fig. 2. **EUCA framework components and creation process.** *Top*: EUCA contains 5 components to support XAI prototyping process. *Bottom*: The range of arrow covers the creation of an EUCA component and its corresponding paper sections. The light blue arrows are preparation stages before the user study, and dark blue arrows indicate the user study phase.

Besides bridging the communication gap between XAI creators and their end-users, EUCA is also a boundary object [3] that bridges the knowledge gap between AI and HCI/UX expertise. Designing end-user-centered XAI is challenging since it requires expertise both in HCI and AI [94]. EUCA is built with a collaborative effort of combining AI and HCI expertise, and XAI creators working in either field can use EUCA to compensate for their missing expertise, or to scaffold the conversation and collaborate more smoothly with team from the other field. For HCI/UX designers who lack constant access to AI experts, EUCA provides tangible design patterns and exemplars that abstract XAI techniques. By having a tool to directly talk to users, EUCA introduces the notion of prototyping and user-centered design to AI researchers and the AI community. It joins the recent effort in XAI field to synergize the HCI and AI community to facilitate interdisciplinary collaboration and communication [11, 94].

In addition to the EUCA framework contribution, the user study also identifies fine-grained end-users' requirements for different explanation goals, such as to calibrate trust, detect bias, resolve disagreement with AI, learn from AI, and to improve the predicted outcomes (Section 6). We also made the EUCA dataset (Supplementary Material S3) available to facilitate modeling end-users' preferences on explanatory forms for personalized and interactive XAI (Section 7.1.3).

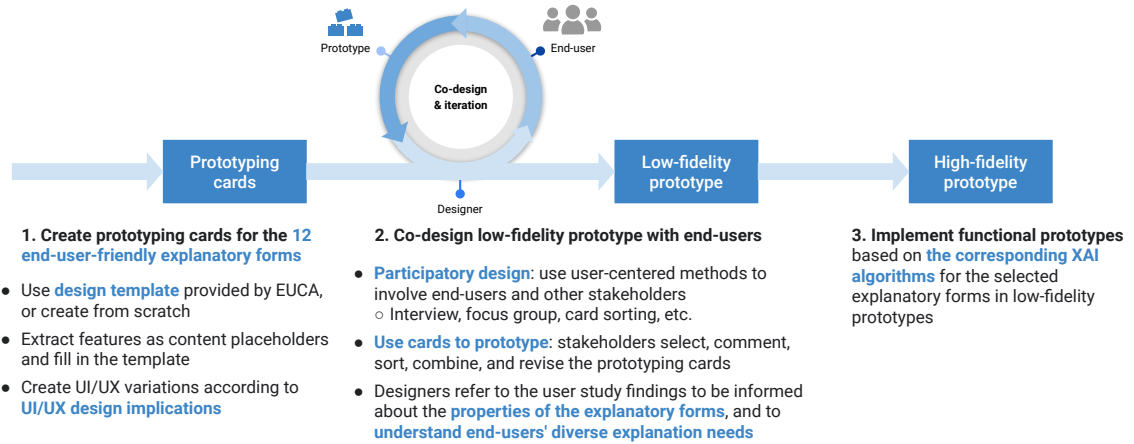

Fig. 3. **The suggested prototyping workflow using EUCA framework.** The blue texts highlight the supporting contents provided by EUCA.

## 2 BACKGROUND AND RELATED WORK

### 2.1 Explainable AI and User-Centered Perspective

Explainable artificial intelligence (XAI), or interpretable machine learning (ML), is usually regarded as a subfield of ML. XAI can be narrowly defined as the information that reveals the decision-making process of an AI model. But a broader definition includes all the necessary background information to make the AI model and its decision-making process transparent and understandable [75], including the training data and AI model performance. We adopt the broader definition of XAI in this paper.

Unlike other ML fields that rarely involve end-users in the technical development and evaluation phase, XAI inherently needs to consider its end-users: Finally, it is the user who will interpret the explanation resulting from an XAI technique. Although in the past few years, the XAI field is booming (largely due to the pervasive use of AI in critical tasks and the legal/ethical requirements on model transparency and accountability [1]), and many XAI techniques have been proposed, most works remain at an algorithmic level. It is unknown whether or how they will work in practice, and what are their suitable use cases. As Lipton criticized, "with a surfeit of hammers, and no agreed-upon nails,"..."we fail to ask what end the proposed interpretability serves" [65]. Although a number of XAI technical taxonomies [75, 103], surveys [42, 91], and technique selection guidance [16] were proposed, they only support the technique selection process in the back-end, not the whole system design including the front-end UI/UX design and user requirements analysis. These technical guidelines also mainly consider the technical aspect rather than usability requirements, and are technical-user-oriented, not end-user-oriented.

The ML and XAI community realized such pitfalls and called for a human-centered perspective and collaboration between HCI and AI fields [65, 72]. The explanation problem is also regaining visibility in the HCI field in recent years. Based on extensive literature analysis, Abdul proposed an HCI research agenda on XAI [11]. Vaughan and Wallach discussed the importance of taking a human-centered strategy when designing and evaluating XAI techniques [94]. Furthermore, XAI is usually part of an AI system, either as an embedded feature or the main component. Therefore, the general human-AI interaction design principles [13, 98] are also applicable to XAI system design.

### 2.2 User-Centered XAI Design Guidance

*2.2.1 Analysis of Prior Works.* To inform the design of user-centered explanations, existing works identify human-centered insights from explanation theories and human-subject studies. We illustrate the comparison of existing user-centered XAI frameworks with EUCA in Fig. 4 and Fig. 5, and state the details below.

Miller summarized the characteristics of explanations from philosophy, psychology, and social science [71]: users prefer simple, *selected* (but may be biased), and *causal* explanations; explanations are *contrastive* to other related predictions, and people tend to seek causal reasoning in a *counterfactual* fashion, i.e.: what would the prediction be if certain features of the input had been different. The explanation is a social process in that humans tailor explanatory contents to different explainability needs and audiences.

Following this line of work, Wang et al. conducted a review on explanation theory literature, and further provided a theory-driven, user-centered XAI framework that describes how the human reasoning process and explanation

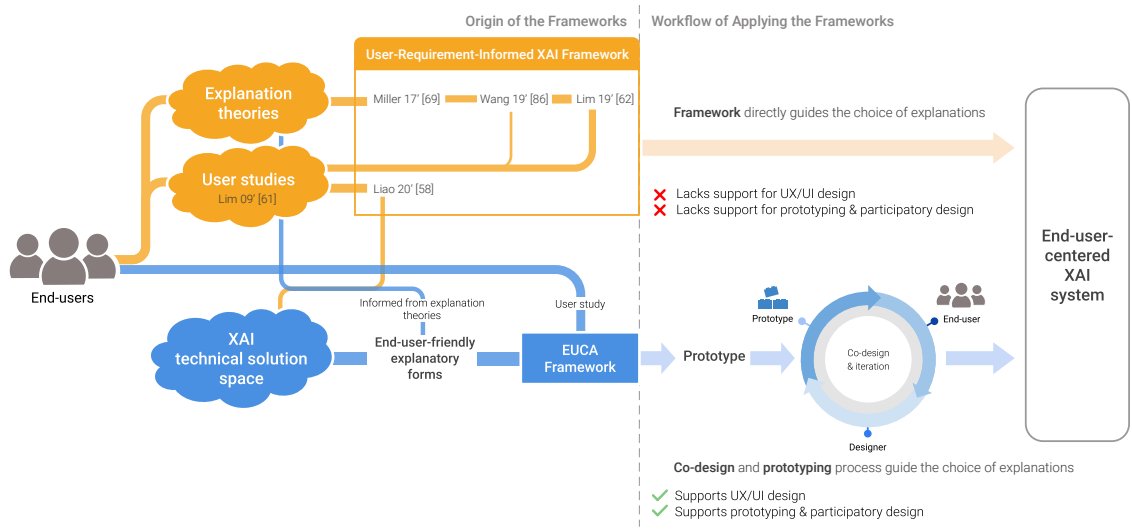

Fig. 4. **Visualizing the distinction between EUCA and prior frameworks regarding their user-centered origin.** [**Left**] Two distinct streams (blue and orange lines) of developing user-centered XAI frameworks: informed by users' requirements only (orange: prior frameworks), and informed by both technical capabilities and users (blue: EUCA). Blue and orange lines link papers with their knowledge sources or prior works. [**Right**] The workflows of using the two types of frameworks in practice. While user-informed frameworks (orange) imply a linear workflow that does not provide opportunities to incorporate users' feedback before implementation, EUCA framework (blue) supports an iterative prototyping process. A detailed comparison regarding the two workflows is expanded in Fig. 5

theories guide explanation system requirements [89]. They suggested the XAI system should support reasoning while mitigating heuristics and bias. Their work is a first attempt in developing user-centered XAI design guidance, but remained at a conceptual and abstract level, and lacked actionable guidance on how to practically implement explanation theories for context-specific tasks and needs.

In their follow-up paper, Lim et al. [64] extended the framework by detailing the explanation types (input, output, certainty, why, why not, what if, how to, and when), and by proposing pathways to link these types to users' three explanation goals: filter causes, generalize and learn, and predict and control. The explanation type taxonomy was first identified by Lim and Dey in 2009, by surveying users' questions in crowdsourcing user studies for context-aware systems [63]. Our explanatory forms overlap with their taxonomy in our category of "supplementary information": input, output, and certainty. As a position paper, their proposed linkage is mainly conceptual and lacks user study evidence. They also did not provide practical guidance or implementation support to illustrate the usefulness of their framework in real-world tasks. In contrast, we included a variety of explanation goals in our user study, and backed our findings on the correlation between explanatory forms and users' needs with quantitative and qualitative user study data.

Liao et al. [60] further explored the idea of providing mapping guidance between users' requirements and explanation types to facilitate human-centered explanation design. The explanation types were identified based on questions users may ask to understand AI. Their framework also provides an additional mapping from explanation types to algorithmic implementation, saving practitioners the effort and needed expertise in identifying the right algorithm to implement. Using the question list and explanation types as a study probe, they further conducted a user study with 20 UX designers to explore the opportunities and challenges of putting XAI techniques into practice. Their results revealed rich details on users' needs for XAI, but failed to show evidences (such as user studies) that the corresponding XAI methods will answer users' questions. Different from EUCA that focuses on the prototyping process, their framework directly guides the choice of explanation types, and it does not provide opportunities to take in users' feedback on the design solutions before the system is implemented.

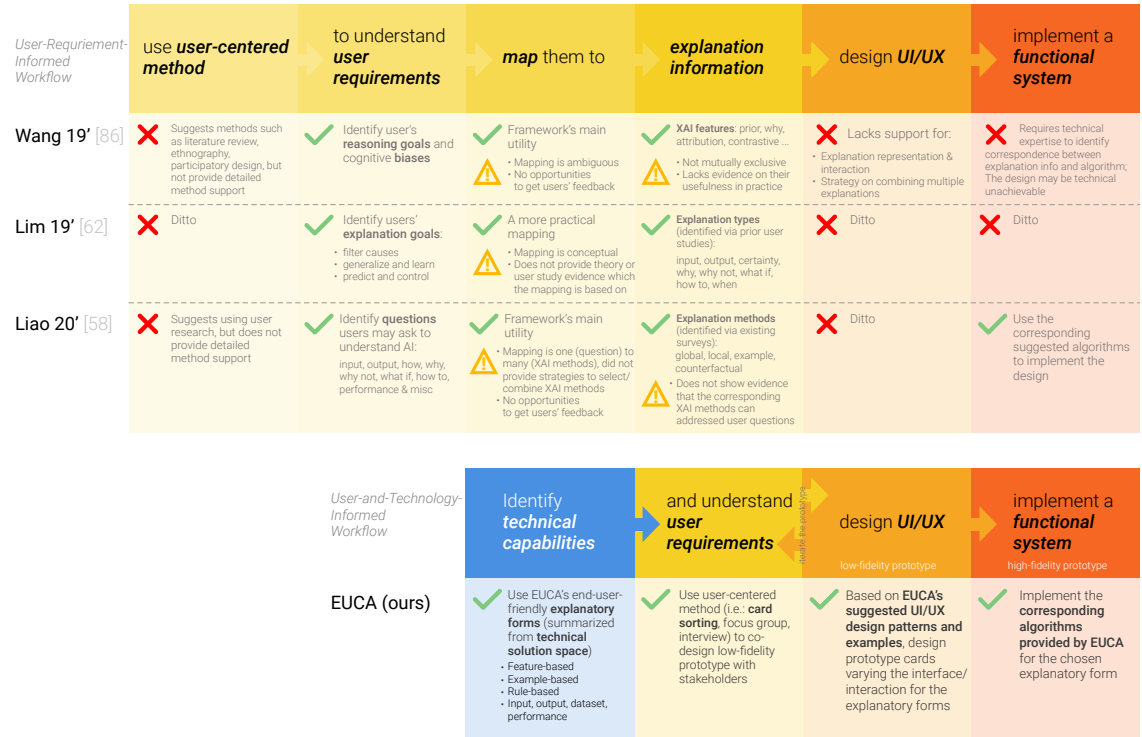

Fig. 5. **Workflow comparison of using two types of framework (top: user-informed; bottom: user-and-technology-informed) in practice.** Prior frameworks imply a user-requirement-informed workflow as shown on the top: user requirement → design → implementation , while EUCA follows a user-and-technology-informed workflow (bottom) by considering technical capabilities . EUCA replaces the direct mapping from user requirements to explanation information, with an iterative prototyping and co-design process (indicated by the back-and-forth arrow in the workflow). For each step in the workflow, we highlight the key actions or results using bold font. ✓ and × indicate whether a step is supported or not supported by a framework, and ⚠ indicates the limitations of applying a framework in a certain step.

*2.2.2 User-Informed vs. User-and-Technology-Informed Paradigm.* While prior frameworks utilized a general user-centered approach by proposing design solutions based on users' requirements, such a direct user-requirement-informed paradigm may not be applicable in the context of XAI, or generally, AI system development. We can abstract the human-centered technology development process as the following workflow:

$$\text{User requirements} \xrightarrow{①} \text{Design} \xrightarrow{②} \text{Implementation} \qquad \text{Workflow (1)}$$

For common technological development, usually the user-centered challenge is in Step ①: the design is directly guided by gathering and analyzing user requirements, which is the focus of previous frameworks. This approach implies that once we find the design solutions, such design can easily be technically implemented (Step ②). This is usually the case for traditional technology development, as the functionality of the system can largely be specified and determined by design, but not the case for AI [98]. Yang illustrated it in a case study on designing an AI-driven clinical decision support system [96], where considering the users' requirements only led to a technically unachievable solution (Step ② is blocked), due to designers' and end-users' limited understanding of AI's current technical capabilities, or lack of training data to train the proposed AI model. In real-world practice, since UX designers usually "know very little about how AI works" [97], to come up with a technical-viable design, UX designers need to work closely with technical teams to incorporate technical solutions in the craft of design, which is the central stage in the AI system design process. Such a relatively novel and unique workflow is the most challenging and unsupported part for design practitioners, and made "working with AI took much longer than when designing other UX products and services", according to Yang et al.'s interview with 13 UX designers who are experienced in AI products [97].

To sum up, the uniqueness of AI and XAI system design requires taking into consideration not only the user side, but fully consider the viable technological solution space [98]. The workflow of designing a user-centered AI or XAI system thus becomes:

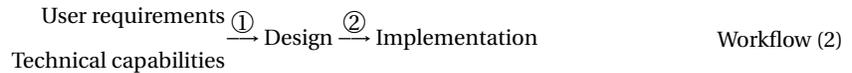

$$\genfrac{}{}{0pt}{}{\text{User requirements}}{\text{Technical capabilities}} \xrightarrow{①} \text{Design} \xrightarrow{②} \text{Implementation} \qquad \text{Workflow (2)}$$

As analyzed in previous Section 2.2.1, prior XAI design frameworks [60, 64, 89] follows user-requirement-informed paradigm (Workflow 1), whereas ours is informed by both end-users and technical capabilities (Workflow 2) (Fig. 4, 5). Although previous frameworks tried to address the unique challenge of technical capability in XAI design, by regulating users' requirements in a pre-defined space (such as pre-defined XAI goals or users' questions), and provided direct mapping from user's requirements to design solutions, such an approach had to compromise design possibilities that reside outside the pre-defined space, and had to limit users' choices, hence may not fully address users' needs. This approach also makes many existing technical solutions under-explored. In contrast, the EUCA framework is a strategic synergy of both XAI techniques and user-centered design methods. To the best of our knowledge, we are the first to take the user-and-technology-informed paradigm to develop an end-user-centered XAI framework.

*2.2.3 Solution-First Design vs. Prototyping.* In addition, prior frameworks implicitly suggested using the framework to directly guide the choice of explanation information. The explanation selection and design processes do not provide opportunities to take in users' feedback until the candidate solution has been designed or implemented. The shortcoming of such a solution-first design approach is that the initial premature solution becomes hard to get over, as the effort in crafting design and implementation becomes an expensive sunk cost [82]. In contrast, prototypes, especially the low-fidelity ones, enable quick and inexpensive trial-and-error, allowing full exploration of the solution space before implementation. We discuss prototyping further in the next section.

**2.3 Prototyping and Co-Design as Necessary Processes towards End-User-Centered XAI**

Given the complex scenarios in real-world design and development practice, using the one-size-fits-all mapping between user's requirements and explanation information provided by existing frameworks may not be applicable.

Prototyping provides a quick-and-dirty way to help XAI system designers understand user-, task-, and scenario-specific requirements, to assess and improve their design. It also sensitizes designers to the scope of AI capabilities, "it is through sketching and prototyping that designers understand what the technology is and can do" [98]. Prior work demonstrated using prototypes and involving stakeholders in the co-design processes in various user-oriented XAI development settings.

To bridge technical XAI with their ambiguous and dynamic real-world use, Wolf [93] proposed to apply scenario-based design [82], an HCI method that mimics the prototyping idea without a tangible prototype [47]. It creates a narrative description to envision the user experience after deployment to guide the system design. Cirqueira et al. demonstrated applying scenario-based requirement elicitation in designing a user-centric XAI system for fraud detection [30].

Similarly, after a literature review, Eiband et al. found there is no consensus in prior works on *what to explain*, and users' demands vary case-by-case [34]. They then presented their six-month participatory design process for a commercial intelligent fitness coach, and demonstrated an iterative prototyping process to answer *how to explain* as follows: in a focus groups workshop with stakeholders, the team members *brainstormed* and *sketched* the UI and user workflow, followed by voting and discussion of the ideas to generate a list of promising implementation ideas. Next they implemented the two most promising ideas as a series of *low- and high-fidelity prototypes*, and evaluated and refined the prototypes in several *user testing*. The process resulted in two high-fidelity prototypes.

Despite prior attempts in incorporating a prototyping process in XAI design, creating XAI prototypes is still a challenging task and requires technical expertise. Practitioners desire support for prototyping tools and methodology, according to a number of user studies with UX designers [60, 96, 97]. Our EUCA framework provides prototyping tools and methodologies to facilitate the creation of low- and high-fidelity prototypes that are technologically feasible. This support is particularly useful for designers who do not have ML expertise or lack constant access to capable data scientists.

## 2.4 End-User-Oriented User Studies on Explanation Information

Existing user studies with non-technical end-user participants were conducted in a case-by-case manner, to understand users' perception of the explanation information and provide insights for XAI design.

Cai et al. [24] examined the effect of *similar examples* and *counterfactual examples* (named comparative explanations in the paper) in a study involving 1150 layperson participants in an online drawing and guessing platform. They found that users who received *similar example* explanations felt they had a better understanding of AI, and perceived AI to have a higher capability. *Counterfactual examples*, however, did not always improve the perceptions of AI as it exposed the limitations of AI and may lead to a confusing or unexpected result.

Narayanan et al. [74] conducted a controlled user study with 600 Amazon Mechanical Turkers to identify how varying different complexities of a *decision set* explanation affect users' ability to interpret it. They found that while almost all types of complexity resulted in a prolonged response time, some types of complexity, such as the number of rules or the number of new features introduced, had a much bigger effect than others, such as repeated features.

While prior works provide individual evidences based on a specific XAI application, our study systematically compares and assesses the strengths, weaknesses, applicable explanation goals, and design implications of each explanatory form in a variety of tasks and explanation goals.

## 3 END-USER-FRIENDLY EXPLANATORY FORMS

Distinct from prior frameworks in which the explanation information is informed from users' requirements only, by applying the EUCA framework, the design of explanation information is informed from both user requirements and technical capabilities. This ensures that the resulting design is technologically achievable. To do so, we began with the observation of existing XAI systems/taxonomies: although the XAI algorithms, models, tasks, and visual representations vary, their resulting explanation information can be abstracted as several recurring forms, such as feature attributions generated by a linear model or algorithms mimicking linear model [12, 67, 79], similar examples from different content-based retrieval algorithms [55], or decision tree and rules. Some forms may be consumed by non-technical users without technical knowledge as a prerequisite. Since the explanatory forms are the final resulting explanation information from the existing XAI algorithms, and the number of explanatory forms is a finite set, we may use the explanatory forms to guide the choice and design of explanations in an XAI system. Because the explanatory forms originate from existing XAI algorithms, once the forms are decided, it is straightforward to implement their associated algorithms, i.e.: [explanatory] form is followed by [the back-end] function (we reverse the famous design maxim "form follows function" [7]). Our approach also echoes the matchmaking design process that identifies potential user domains ("nails") for numerous existing XAI techniques ("hammers") [20].

Based on the above insights, we explored the XAI solution space by extracting the resulting explanation information from existing technical literature in AI, HCI, and information visualization fields via literature review, then selected and summarized end-user-friendly explanatory forms based on the following criteria:

(1) The explanatory forms must be end-user-friendly, i.e., users are not required to have background knowledge in AI or machine learning techniques to understand the explanation.
(2) The explanatory forms are generally mutually exclusive regarding their underlying information. We noted sometimes the explanation information can be attributed to different XAI types and concepts which are entangled with each other, e.g.: *causal* explanations may be expressed as feature attributions or rules; *counterfacutal* explanations can be represented as counterfactual features, examples, or rules; feature attribution can be *global* as well as *local*. We selected forms that are mutually exclusive, so that they can act as building blocks that represent the elemental explanation information, and their combination would not be redundant/repeated in an XAI system.

The literature review process and the list of surveyed literature are detailed in Supplementary Material S1. We ended up with 8 explanatory forms in three categories: explaining using features, examples, and rules. In addition, we added necessary supplementary information to make the explanation complete, including input, output, dataset, and performance. A total of 12 end-user-friendly explanatory forms are included in the EUCA framework (Fig. 1).

The end-user-friendly explanatory forms are a familiar and common language to both end-users and XAI practitioners, so that they can facilitate user requirements communication and the co-design process. Since the forms are summarized from a technically achievable solution space and shown as UI design patterns, it also bridges the expertise gap between HCI/UX designers and AI developers. Next, we introduce each explanatory form, and summarize their possible visual representations from the surveyed literature to facilitate UI/UX design. Table 1 highlights the key properties of the 12 explanatory forms, and Fig. 1 shows visual examples.

Table 1. **The End-User-Friendly Explanatory Forms.** We indicate whether a form is a global (explaining the model's overall behavior), or local explanation (explaining a decision for an individual instance). We also give its applicable input data types: Tabular - tabular data, Img - spatial-structured data (e.g., image, graph), Txt - sequential data (e.g., text, signal). The ⋆ indicates the user-friendly level based on the card selection frequency in the user study (1: least friendly, 3: most friendly). Its pros, cons, design implications, and applicable explanation goals are the key findings of RQ1 from the user study, followed by its associated algorithms for implementation.

| Explanation Category | Explanatory Form | Visual Representations | Pros | Cons | UI/UX Design Implications | Applicable Explanation Goals | XAI Algorithm Examples |
|---|---|---|---|---|---|---|---|
| Feature-based explanation | **Feature Attribution** Local/Global Tabular/Img/Txt ⋆⋆⋆ | Saliency map; Bar chart | Simple and easy to understand; Can answer *how* and *why* AI reaches its decisions. | Illusion of causality, confirmation bias | Alarm users about causality illusion; Allow users to set thresholds on feature importance score, and show details on-demand | To verify AI's decision | LIME [79], SHAP [67], CAM [105], LRP [17], TCAV [53] |
| | **Feature Shape** Global Tabular ⋆⋆ | Line plot | Graphical representation, easy to understand the relationship between one feature and prediction | Lacks feature interaction; Information overload if multiple feature shapes are presented | Users can inspect the plot of their interested features; May indicate the position of local data points (usually users' input data) | To control and improve the outcome; To reveal bias | PDP [35], ALE [15], GAM [87], SHAP dependence [67] |
| | **Feature Interaction** Global Tabular ⋆ | 2D or 3D heatmap | Show feature-feature interaction | The diagram on multiple features is difficult to interpret | Users may select their interested feature pairs and check feature interactions; or XAI system can prioritize significant feature interactions | To control and improve the outcome | PDP [35], ALE [15], GA2M [25], SHAP interaction [67] |

| Explanation Category | Explanatory Form | Visual Representations | Pros | Cons | UI/UX Design Implications | Applicable Explanation Goals | XAI Algorithm Examples |
|---|---|---|---|---|---|---|---|
| Example-based explanation | **Similar Example** Local Tabular/Img/Txt ★★★ | Data instances as examples | Easy to comprehend, users intuitively verify AI's decision using analogical reasoning on similar examples | It does not highlight features within examples to enable users' side-by-side comparison | Support side-by-side feature-based comparison among examples | To verify the decision | Nearest neighbour, CBR [55] |
| | **Typical Example** Local/Global Tabular/Img/Txt ★★ | Data instances as examples | Use prototypical instances to show learned representation; Reveal potential problems of the model | Users may not appreciate the idea of typical cases | May show within-class variations; or edge cases | To verify the decision; To reveal bias | k-Mediods, MMD-critic [52], Generate prototype [68, 84], CNN prototype [27, 59] |
| | **Counterfactual Example** Local Tabular/Img/Txt ★★ | Two counterfactual data instances with their highlighted contrastive features, or a progressive transition between the two | Helpful to identify the differences between the current outcome and another contrastive outcome | Hard to understand, may cause confusion | User can define the predicted outcome to be contrasted with, receive personalized counterfactual constraints; May show controllable features only | To differentiate between similar instances; To control and improve the outcome | Inverse classification [58]), MMD-critic [52], Progression [51], Visual [38] |

| Explanation Category | Explanatory Form | Visual Representations | Pros | Cons | UI/UX Design Implications | Applicable Explanation Goals | XAI Algorithm Examples |
|---|---|---|---|---|---|---|---|
| **Rule-based explanation** | **Decision Rules/Sets** Global Tabular/Img/Txt ★★ | Present rules as text, table, or matrix | Present decision logic, *"like human explanation"* | Need to carefully balance between completeness and simplicity of explanation | Trim rules and show on-demand; Highlight local rule clauses related to user's interested instances | Facilitate users' learning, report generation, and communication with other stakeholders | Bayesian Rule Lists [95], LORE [41], Anchors [80] |
|  | **Decision tree** Global Tabular/Img/Txt ★★★ | Tree diagram | Show decision process, explain the differences | Too much information, complicated to understand | Trim the tree and show on-demand; Support highlighting branches for user's interested instances | Comparison; Counterfactual reasoning | Model distillation [36], Disentangle CNN [102] |

**3.1 Feature-Based Explanation**

Feature-based explanations are the most common form of explanation information. A *feature* is a piece of information that describes the input data. It could be the raw representation of the input (such as image pixels, sound wave signals), or descriptive characteristics of the input summarized/designed by the human (such as house features presented as tabular data), or automatically learned by AI. For example, a real estate agent can describe a house by its size, location, and age, three descriptive features; The feature of an image can be each individual pixel, or a group of pixels highlighting the object of a car, or the abstract concept of "car". To use features in an explanation, the feature representation must be human-interpretable.

Features are the elements of explanation. Features to an explanatory form is like words to a sentence. Different explanatory forms (sentence types) describe the relationship between features (words) and outcomes using their individual logical forms: feature-based explanations describe features $A_i, A_j, A_k, \ldots$ and their relationship with the outcome $Y$ in a quantitative manner; example-based explanations present features within its context as instances $X_i, X_j, \ldots$; and rule-based explanations organize features and outcomes by logic and conditional statements.

The feature-based explanations consist of three explanatory forms:

*3.1.1* **Feature Attribution**. It indicates which features are important for the decision, and what are their attributions to the prediction: $p(Y = y | A_i = a)$. For example, it can be a list of key features and their importance scores to the house price prediction, or a color map overlaid on the input image indicating the important parts/objects for image recognition.

*Visual representation*: Its visual representations largely depend on the feature data type. For image and text data, overlaying a **saliency map** or color map on the input is the most common visualization. It uses sequential colors to code the fine-grained feature importance score for each individual feature (could be a pixel for image input, a word for text data). For image/video input data, other popular visualizations include using *segmentation masks* or *bounding boxes* on important image objects/parts.

To visualize multiple feature attributions for tabular or text data, a **bar chart** is a typical choice. Variations of the bar chart include waterfall plot, treemap, wrapped bars, packed bars, piled bars, Zvinca plots, and tornado plot. Compared to a bar chart that shows a point estimation of feature importance, a *box plot* can be used to visualize the probabilistic distribution of the feature importance score. Its variations include violin plot and swarm plot that show more detailed data distribution and skewness.

*3.1.2* **Feature Shape**. It describes the relationship between one particular feature $A_i$ and the outcome $Y$: $p(Y|A_i)$, such as the house size to the predicted house price.

*Visual representation*: For a continuous feature (such as height, temperature, i.e.: measurement on a scale), a **line chart** is the most common visualization, depicting whether the relationship between the feature and outcome is monotonic, linear, or more complex. The line chart can be accompanied by a scatter plot detailing the position of individual data points.

For a categorical feature (such as gender, season), a *bar chart* can be used.

*3.1.3* **Feature Interaction**. When features interact with each other, their combined effect on the outcome may not be a linear summation of each feature's individual effect. Feature interaction considers such an interdependence effect, and shows the total interaction effect of multiple features $A_i, A_j$ on the outcome $Y$: $p(Y|A_i, A_j)$. It can be regarded as an extension of feature shape by taking multiple features (instead of one feature) into account.

*Visual representation*: **2D or 3D heatmap** is usually used to visualize the combined effect of feature interactions on prediction. Limited by the visualization, a heatmap can show feature interactions for at most three features (using 3D heatmap). More complicated multiple paired feature-feature interactions can be visualized using matrix heatmap, node-link network, or contingency wheel.

### 3.2 Example-Based Explanation

Human uses examples to learn and explain. Examples carry contextual information and are intuitive for end-users to interpret. Three different types of examples are included: similar, typical, and counterfactual examples.

*3.2.1* ***Similar Example***. Similar examples are instances $X_i, X_j, \ldots$ that share similar features with the query input $X$ regardless of their predictions. Mathematically, they are instances that have minimal distance to the query: $\operatorname{argmin}_{X_i \in S} d(X_i, X)$, where $S$ is the set of training data or the whole input space, and $d$ is the distance or dissimilarity measure in the input space. For example, for a house to sell, its similar examples can be houses in the adjacent area with similar features such as house size, age, etc.

The differences among similar, typical, and counterfactual examples are listed in Table 2: For a similar example, although it shares *similar features* with the query instance, their predictions may be *the same or different*. Whereas for a counterfactual example, it not only shares *similar features* with the query instance, but also has a *different prediction*. For a typical example, it has *the same prediction* as the query instance, regardless of their features.

| **Explanatory Form** | **Features** | **Prediction** |
|---|---|---|
| **Similar** Example | similar | the same or different |
| **Typical** Example | similar or different | the same |
| **Counterfactual** Example | similar | different |

Table 2. **Distinctions among the three example-based explanatory forms** by comparing their features and prediction with the query instance.

*3.2.2* ***Typical Example***. A typical or prototypical example is a representative instance of a certain prediction, i.e.: an instance $X_i$ that maximizes the probability of the target prediction $t$: $\operatorname{argmax}_{X_i \in S} p(f(X_i) = t)$. For example, a typical example of the diabetes prediction could be a patient who exhibits typical characteristics (such as a high blood sugar level, an abnormal hemoglobin A1C level) that could be diagnosed as diabetes.

*Visual representation*: For similar and typical examples, it is straightforward to show several examples with their corresponding predictions.

*3.2.3* ***Counterfactual Example***. For a counterfactual example $X_i$, its features are similar to the input $X$, but has minimal feature changes so that its outcome $t'$ is distinct from the outcome $t$ of the original input $X$, i.e.: $\operatorname{argmin}_{X_i \in S} d(X_i, X), s.t. \max p(f(X_i) = t')), t' \neq t$. For example, given an input $X$ that is predicted to have diabetes $t$, its counterfactual example $X_i$ is an instance that is predicted to be healthy $t'$, with the same or similar features to those of $X$, except that its blood sugar level is lower than $X$.

We noted that counterfactual explanations can also be expressed as counterfactual features or rules. However, a counterfactual feature/rule can not be a standalone explanation in an XAI system, they must reside within a certain context by assuming all other features are constant. To make the explanation information complete, we include the counterfactual explanation in the form of example.

*Visual representation*: Counterfactual examples can be shown as two instances and their predictions, with their **counterfactual/contrasting features highlighted**, or a **transition** from one instance to the other by gradually changing the counterfactual features.

### 3.3 Rule-Based Explanation

Rule-based explanations are explanations where the decisions of the model, in whole or in part, can be described succinctly by a set of logical IF-THEN statements, mimicking human reasoning and decision making. It also implies the decision boundary, and that may be convenient for counterfactual reasoning. The rule-based explanation is a global explanation of the model's overall behavior. It includes the following two explanatory forms: decision rules and decision tree. We note decision rule and decision tree actually carry similar explanation information. However, since they are usually generated by different XAI algorithms, and their representation formats (text vs. diagram) are different to users, we included them as two separate explanatory forms.

*3.3.1* **Rule**. The decision rules or decision sets are simple IF-THEN statements with conditions and predictions. For example: IF blood sugar is high, AND body weight is overweight, THEN the estimated diabetes risk is over 80%.

*Visual representation*: Rules are usually represented using **text**. Other representing formats include table [26] or matrix [73] to align, read, and compare rule clauses more easily.

*3.3.2* **Decision Tree**. Decision tree represents rules graphically using a tree structure, with branches representing the decision pathways, and leaves representing the predicted outcomes.

*Visual representation*: The most common representation is to use a node-link **tree diagram**. Other visual representations to show the hierarchical structure include treemap, cladogram, hyperbolic tree, dendrogram, and flow chart.

### 3.4 Supplementary Information of an XAI System

In addition to the above explanatory forms generated by XAI algorithms, an XAI system needs to present some necessary background or supplementary information to end-users, such as input, output, decision confidence/certainty, model performance metrics, training dataset information such as data distribution, etc. We included the following essential and common ones in our framework, and indicates whether each is a global (explaining the model's overall behavior) or a local explanation (explaining the decision on an individual instance):

(1) **Input**, **output** (local): input is end-users' input data, and output is AI's prediction on the given input.
(2) **Certainty** (local): Since the prediction from AI models is usually probabilistic, the certainty score shows the case-specific certainty level about how confident the model is in making this particular decision.
(3) **Performance** (global): Performance metrics (such as model accuracy, confusion matrix, ROC curve, mean squared error) help end-users to judge the overall decision quality of the model, and to set a proper expectation on the model's capability, as suggested in the human-AI interaction guideline [13].
(4) **Dataset** (global): It describes the information about the dataset which the AI model is trained on, such as the training data distribution. It may help end-users to understand the model and identify potential flaws or biases in the dataset.

In this section, we introduced the twelve end-user-friendly explanatory forms from the technical domain. They provide feature-, example-, and rule-based explanations. Next, we will examine usability of these explanatory forms by conducting a user study with end-users.

## 4 USER STUDY METHOD

We conducted a user study with 32 layperson participants. The user study utilized an interview and card selection & sorting methodology. It aims to elicit the process of using EUCA prototyping workflow to design XAI low-fidelity prototypes in different AI-assisted critical decision-making tasks. The primary goal of the user study is to understand end-users' perception of the explanatory forms. Specifically, we aim to identify the properties of explanatory forms to incorporate them as design support into the EUCA framework. Their properties include: strengths, weaknesses, applicable explanation goals, and UI/UX design implications. The secondary goal is to use EUCA as a study probe to understand end-users' various requirements for different explanation goals, such as to calibrate trust, detect potential biases, verify the decisions when user disagrees with AI, and to learn from AI, etc. (Section 4.2.2). Therefore, our research questions (RQ) are:

**RQ1**: For each end-user-friendly explanatory form, what are its strengths, weaknesses, applicable explanation goals, and UI/UX design implications?

**RQ2**: What are end-users' requirements for different explanation goals?

### 4.1 Participants and Recruitment

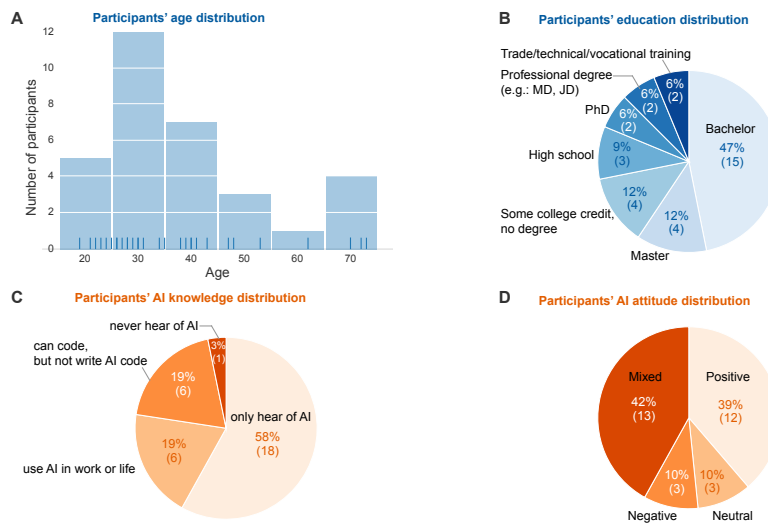

Fig. 6. **Participants' demographic information**. (**A**) Histogram of participants' age distribution. The sticks on the horizontal axis show each participant's age. (**B**) Pie chart on participants' educational level. Numbers in parentheses represent the number of participants in that category. (**C**) Pie chart on participants' familiarity with AI. (**D**) Pie chart on participants' attitudes towards AI; Positive attitudes include *"interested"* in and *"excited"* to use AI; Negative attitudes consist of *"skeptical"* and *"concerned"* about AI; A mixed attitude means participants hold both positive and negative attitudes towards AI.

We recruited layperson participants via a convenience sampling method by advertising posters at public libraries, community centers, and online community boards in the Greater Vancouver area over a 3-month period in 2019. The inclusion criteria were: 1) adult (19 years old and above); and 2) do *not* have prior technical knowledge in machine learning, data science, or artificial intelligence. A total of 32 participants were enrolled in the study (Female = 16; Age: 38.2±16.0, range 19-73). Participants' occupations covered a variety of industries e.g.: technology, design, car insurance, finance, psychology, construction, sales, food/cooking, law, healthcare, government/social services, and retired. For participants who use AI in work or life (6 participants, 19%), they used AI software such as Google Assistant to play music, navigate traffic, chat with clients, and help drive investment decisions. Figure 6 shows the distribution of participants' age, educational background, familiarity with and attitudes towards AI. Participants' detailed demographics are in Supplementary Material S2. The participants were thanked with $25 CAD for their time and effort in the study. The study is approved by Simon Fraser University Research Ethics Board (Ethics number: 2019s0244).

### 4.2 The Interview Instrument

*4.2.1* **Critical Decision-Making Tasks**. We focus the scope of the study on AI-assisted critical decision-support tasks, where explanations have high utility as shown in previous research [23, 32, 62], and AI could not be delegated to have full automation because of the high-stakes nature of the tasks and the liability issue. In this study phase, we did not included domain experts. Therefore, we deliberately designed the tasks so that decisions can be made based on common sense without requiring domain knowledge. We designed four decision-making tasks reflecting the diversity of AI-supported critical decision-making. They are:

(1) **House** task: users use AI to get a proper estimate of their house price.
(2) **Health** task: users use AI to predict his/her diabetes risk.
(3) **Car** task: users decide whether to buy an autonomous driving vehicle.
(4) **Bird** task: users use AI bird recognition tool to prepare for an important biology exam.

The four tasks are critical decision-making scenarios, because their decisions have significant consequences on one's health and life (Health and Car Task), finance (House Task), or education (Bird Task). The four tasks covered common input data types of tabular, sequential, image and video data. The corresponding datasets of the four tasks are publicly available (Table 3), so that the resultant low-fidelity prototypes from this user study can be actualized as high-fidelity functional prototypes for task-specific studies in future work.

*4.2.2* **End-Users' Explanation Goals**. Even for the same user and task, end-users' explanation goals, i.e.: the trigger point or motivation to check the explanation of an AI system, may vary in different contexts or usage scenarios. In our study, we aim to capture the fine-grained details of end-users' requirements for different explanation goals. We summarize the following potential *explanation goals* from prior works [32, 40, 64, 71, 75, 81] as follows:

- **Calibrate trust**: trust is key to establish human-AI decision-making partnership. Since users can easily distrust or overtrust AI, it is important to calibrate trust to reflect the capabilities of AI systems [88, 104].
- **Ensure safety**: users need to ensure safety of the decision consequences [32].
- **Detect bias**: users need to ensure the decision is impartial and unbiased [64, 75].
- **Unexpected prediction**: the AI prediction is unexpected, and/or users disagree with AI's prediction [40].
- **Expected prediction**: AI's prediction aligns with users' expectations [40].

| *AI-Assisted Critical Decision-Making Tasks* | | **HOUSE TASK**<br>Sell a house<br>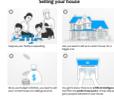 | **HEALTH TASK**<br>Check diabetes risk<br>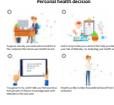 | **CAR TASK**<br>Buy a self-driving car<br>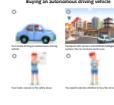 | **BIRD TASK**<br>Prepare for exam<br>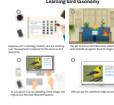 |
|---|---|---|---|---|---|
| *Explanation Goals* | **Calibrate Trust** | You doubt whether to trust the AI tool or not | You doubt whether to trust the software prediction on your diabetes risk | n/a | You don't know whether to trust the results from the website or not |
| | **Ensure Safety** | n/a | n/a | You need to know whether the autopilot mode is safe and reliable | n/a |
| | **Detect Bias** | n/a | You doubt whether the software will perform the same among people with different gender, age, or ethnicity group | You want to know if the autopilot mode performs robustly under varying road, weather, and light conditions. | n/a |
| | **Unexpected Prediction: Disagreement with AI** | AI's prediction aligns/does not align with your own estimation | You maintain good health with no major diseases or a family history of diabetes/Diabetes tends to run in your family, and you're afraid of getting it someday, and AI predicts your chance of getting diabetes is low/high | You notice the car sometimes drives much slower than the expected speed limit | The results sometimes do not align with your knowledge |
| | **Differentiation Similar Instances** | n/a | n/a | n/a | In the exam, you need to write a short statement to differentiate different birds |
| | **Learn from AI** | n/a | n/a | n/a | Is it a good tool to improve your learning and help you know more about bird taxonomy? |
| | **Improve the Predicted Outcome** | You need to decide whether to do a renovation or replacement of appliances to increase your house value, and which action is the most cost-effective | You want to know how to adjust your lifestyle accordingly to lower the risk of diabetes | n/a | n/a |
| | **Communicate with Stakeholders** | You need to communicate your decision with your family | You need to need to inform family members and consult your doctor | You need to communicate with your family about your judgment on the car's safety | n/a |
| | **Generate Reports** | n/a | n/a | n/a | In the exam, you need to write a short statement on how you recognize the bird as such a species |
| | **Multi-Objectives Trade-Off** | n/a | You're aware that the insurance company may use such a prediction from the software to determine your insurance premium and benefits | You're easy to get motion sickness, and you notice you seem to get car sick more frequently in autopilot mode | n/a |
| *ML problem type* | | Regression | Regression | Classification | Classification |
| *Input data type* | | Tabular data | Tabular/sequential data | Image/video data | Image data |
| *Available dataset* | | Boston housing [10] | Diabetes dataset [6] | BDD100K [100] | CUB-200 dataset [92] |

Table 3. **The four tasks and their explanation goals used in the interview**.

- **Differentiate similar instances**: due to the consequences of wrong decisions, users sometimes need to discern similar instances or outcomes. For example, a doctor differentiates whether the diagnosis is a benign or malignant tumor [64].
- **Learn from AI**: users need to gain knowledge, improve their problem-solving skills, and discover new knowledge [32, 40, 64, 75].
- **Improve the predicted outcome**: users seek causal factors to control and improve the predicted outcome [64, 71, 75].
- **Communicate with stakeholders**: many critical decision-making processes involve multiple stakeholders, and users need to discuss the decision with them [71].

- **Generate reports**: users need to utilize the explanations to perform particular tasks such as report production. For example, a radiologist generates a medical report on a patient's X-ray image [40].
- **Multiple objectives trade-off**: AI may be optimized on an incomplete objective while users seek to fulfill multiple objectives in real-world applications. For example, a doctor needs to ensure a treatment plan is effective as well as having acceptable patient adherence. Ethical and legal requirements may also be regarded as objectives [32].

Each task is accompanied by several explanation goals as shown in Table 3. The tasks and explanation goals were presented in the form of storyboards using graphics and text (Fig. 7). Supplementary Material S2 details the interview schedule and materials.

*4.2.3 Creating Prototyping Cards from Explanatory Forms.* To understand end-users' perceptions of the 12 explanatory forms, we need to instantiate them as low-fidelity prototyping cards for each task. We illustrate this process below:

(1) **Create prototyping card templates** We started by creating templates that visualize the explanatory forms. We selected the most common visualizations, based on the summarized visual representation from previous literature in Section 3. For example, we used bar chart and color map to visualize feature attribution for tabular and image data respectively. Each individual card shows one explanatory form. For some explanatory forms (such as feature attribution and counterfactual example), we created multiple cards with different variations of their visual representations.
(2) **Extract features as content placeholder** We then manually extracted several interpretable features given the AI task. For instance, in the house price prediction task, we extracted features of house size, age, etc. In the self-driving car task, we extracted saliency objects such as traffic signs, road markers, cars, and pedestrians. As a quick prototyping, the feature content may not necessarily reflect the real content generated by XAI algorithms. They mainly serve as content placeholders.
(3) **Fill the prototyping templates with content placeholder** The extracted features were then used to fill in the prototyping card templates. The final prototyping cards are shown in Figure 1 and Supplementary Material S2.

After interviewing the first five participants, we revised some prototyping cards based on participants' feedback. For instance, we indicated the position of the input data point on the feature shape and feature interaction cards. We also removed several variations of the cards (such as using a table to represent rules) since participants found them difficult to interpret.

## 4.3 Study Procedure

The study session consisted of a one-to-one, in-person, open-ended, semi-structured interview and a card selection & sorting.

*4.3.1 Interview.* The interview consisted of two rounds (Fig. 7). **Round 1** is to familiarize participants with the tasks and explanation goals, and to understand end-users' explanation goals for XAI (**RQ2**) before showing them prototyping cards of the explanatory forms. The participant was first introduced to an AI-assisted decision-making task and its corresponding explanation goals. Each task and explanation goal were shown as storyboards color-printed on paper. For each explanation goal, we asked the participants whether they accept the AI as decision-support, and need the AI to explain its decision. If explanations were needed, we then asked what explanations/further information they request.

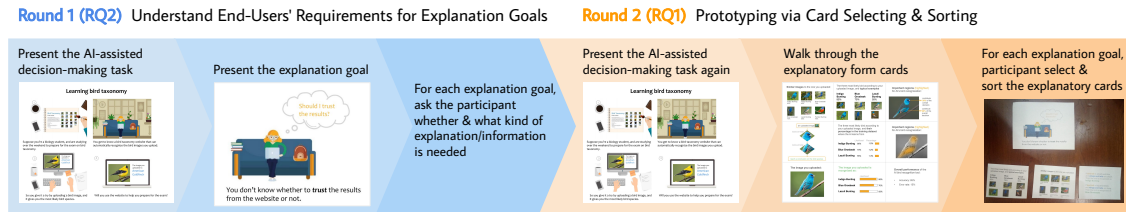

Fig. 7. **The user study procedure**. The study consisted of two rounds corresponding to the two research questions. In the above example, we use the Bird task and the explanation goal of calibrating trust.

After discussing all explanation goals for one task, the participant entered **Round 2**: card selection & sorting, which is detailed in the next section. At the end of the interview, the participants filled out a demographic questionnaire. The mean and median study duration were both 67 minutes (each participant's specific data are detailed in Supplementary Material S2). We audio-recorded the interviews, made observational notes on the card selection & sorting process, and took pictures of the card selection & sorting results.

*4.3.2  Prototyping via Card Selection and Sorting.* For each decision-making task, the participant first revisited the task. Then the researcher walked him/her through the prototyping cards that visualize the explanatory form. In this process, the participant could ask questions if s/he did not understand or needed clarification. S/he could also comment on each card. Before moving on to the next step, we asked the participant and made sure they had no questions on these cards.

Next, for each explanation goal, the researcher asked the participant to select, rank, and combine the prototyping cards which they found were the most useful ones and could meet the current explanation goal. Participant could also modify the existing cards, or sketch on blank cards to create new prototyping cards, and add the sketched cards to the card sorting. After sorting the cards, participants were asked to comment on why they selected or did not select a card, and their rationals for making such a sorting (**RQ1**). After the card selection & sorting, they were asked whether the combination of cards would fulfill the explanation goals.

### 4.4  Data Analysis

We utilized a mixed method to analyze the data. For qualitative analysis, we analyzed the interview data using an inductive thematic analysis approach [22]. A total of 2175 minutes of interviews were recorded and transcribed. We performed coding using Nvivo software. Three members of the research team started with an open coding pass to individually create a list of potential codes. Two additional sets of codes were also applied: 1) the 11 explanatory *explanation goals* (listed in Section 4.2.2); 2) the 12 *explanatory forms* (listed in Section 3). Upon discussion and applying the affinity diagram process, a unified coding scheme was devised. The coding scheme was not task-specific, as we aim to identify and abstract the general themes despite their different tasks. Two team members independently coded one transcription using this scheme. The first pass of inter-rater reliability Kappa score was 0.43. After an in-depth discussion with the research team, we further clarified the code definition, merged potential overlapping codes, and removed redundant codes in the coding scheme. The second pass of inter-rater reliability Kappa score was 0.88 on two more transcriptions. The first author analyzed all interview transcripts twice, and the other coder analyzed half of the transcripts.

We also conducted quantitative analysis on card selection & sorting, and on participants' responses to the questions asked in the interview. Since the quantitative results are less relevant to the main research questions, we put the quantitative methods and results in Supplementary Material S2.

### 4.5 Presentation of Results

The *explanation goals* are marked blue, and *explanatory forms* are marked orange. We highlighted key messages using **bold** font. Whenever necessary, we included participants' verbatim quotes despite some minor grammatical errors. Some quotes had their task and explanation goal listed in the parentheses.

## 5 RQ1: PROPERTIES OF THE END-USER-FRIENDLY EXPLANATORY FORMS

For each explanatory form, the user study (at Round 2) identifies their properties including pros, cons, applicable explanation goals, and design implications. The key findings are summarized in Table 1. Those properties provide design insights from an end-users perspective to facilitate *practitioners*' prototyping process. The weakness and design implication part can also inspire *XAI researchers* to improve existing XAI methods/interfaces or propose new ones.

### 5.1 Feature Attribution

*5.1.1 Pros.* In the study, we used a bar chart to represent feature attribution or importance score for tabular data, and color map and bounding box object detection for image data (Fig. 1). All participants **intuitively understood feature attribution**, and over half selected it (143/248) and ranked it at a relatively top position.

> *"Feature attribution uses a simple way to highlight the most important parts, and you can see very clearly at your first sight how this can be recognized."* (P04, Bird, Learning)
> *"It's easy to read. ...And you have a bar (chart) here it's really clear information that people understand instantly."* (P28, House, Trust)

By showing *"finer details"* (P10) and *"breakdown and weights of features"* (P23) *"that AI took into account"* (P31), participants perceived feature attribution **can answer *"how"* and *"why"* questions**.

> *"tells me why"* (P20), *"gives me the behind the scenes"* (P24), *"tells me how AI read things and how it makes decisions"* (P03), *"have an understanding of how much weight AI is giving to each of the factors"* (P22), and *"identify key aspect, ...support its reasoning"* (P18).

*5.1.2 Applicable explanation goals.* By checking the ranking of feature importance score, participants would instantly *"compare with my own judgment, to see if that aligns with my feature attribution"* (P01, Car, Safety), especially when participants need to **verify AI's decision**.

*5.1.3 Cons.* Although a causal relationship may not be confirmed, some participants tended to **assume feature attribution is causal**, or simplify the relationship among features by assuming they are **independent of each other**. This was usually occurred when they were seeking explanation to improve the predicted outcome. Furthermore, participants were likely to be informed by the feature importance score to prioritize the most important features to take actions upon.

> *"Seeing that body weight is more important than exercise, I think I will focus on changing what I ate, instead of like responding by going to the gym everyday."* (P16, Health, Improvement) – Relies on feature attribution to improve the outcome.
>
> *"It (feature attribution) shows what are the most important factors that AI has taken into account, so you could target the biggest factors."* (P31, Health, Improvement) – Assumes a causal relationship and prioritizes her action accordingly.
>
> *"If my blood sugar puts me at a super high risk here, but my caloric intake doesn't actually put me at that higher risk, it's like a lower risk, then I would rather just focus on blood sugar. "* (P22, Health task) – Ignores the complex interaction between blood sugar level and caloric intake.

*5.1.4* **Design Implications**. To avoid the above causal illusion [69], UI/UX design may need to **alarm users** (explicitly or implicitly) that changing the important features may not necessarily lead to the outcome change in the real world, because most AI models can only capture the correlation between features and prediction, and correlation does not necessarily imply causality.

For the prototyping card design, designers may consider **varying different UI/UX representations** of the feature importance, such as showing the feature ranking only and allowing users to check the detailed attribution scores on demand, or allowing users to set a threshold on the attribution score and only showing features above the cut-off value, as suggested by a few participants.

> *"If the percentage (of the feature) is below the cutoff value, the users does not need to see (the feature), reduce the cognitive load."* (P04, Bird, Learning)

**5.2 Feature Shape**

*5.2.1* **Pros**. Participants **liked its graphical representation** of showing the relationship between one feature and prediction.

> *"It (feature shape on exercise and diabetes risk) feels so easy to latch onto like it's something that you can impact and something that's very tangible."* (P22, Health, Trust)

*5.2.2* **Applicable Explanation Goals**. The slope of the curve in feature shape line chart allows users to easily check how changing one feature would lead to the change of the outcome. Thus, many participants intuitively used feature shape for **counterfactual reasoning**, especially to improve the predicted outcome.

> *"I would be interested to see how much like here (feature shape) increasing the exercise by a small amount actually makes a really big difference. So that's also helpful to decide what you should be focusing on to try to avoid it (diabetes). The shape of the curve actually helps. Coz if I was out here [pointing to the flat part of the curve], then it would not be as helpful for me to increase my exercise."* (P16, Health, Improvement)

By showing the relationship between the protected feature (such as gender, ethnicity, [70]) and outcomes (such as loan approval), the feature shape explanation is also helpful to reveal potential bias, i.e.: to check if different assignments of the protected features (male, female) will lead to prediction differences.

> *"If these features are related to diabetes, then it (AI) should present some (feature shape) cards to tell me if the gender, age and ethnicity (will affect diabetes prediction), so this image (feature shape) would be really helpful."* (P02, Health, Bias)

*5.2.3* **Cons**. One drawback of feature shape pointed out by a few participants is that it does **not considerate feature interactions**.

> *"This one (feature shape on house size and price) is not based on the bigger the house, the higher you can sell, because it is based on a lot of features. Let's say the house is 2000 square feet. It was built in 1980. Another one is 1000 square feet, but it's just built a decade ago. So its (the latter) price will be much higher than this one (the former). You cannot just base on a house area and then determine the price."* (P30)

Another drawback is, since a feature shape plot can only explain one feature at a time, to present explanation for multiple features, the interface will be occupied by multiple feature shape plots which may –

> *"make your page so **overloaded**, so people just get tired. You want to make it as clear as possible. So if (there is) some unnecessary information people just intimidated."* (P28)

*5.2.4* **Design Implications**. One suggestion for the above weaknesses is that feature shape can be accompanied by other explanatory forms and **show on-demand**. Users can select their interested features from a feature list, or click a feature from other explanatory forms (such as feature attribution, counterfactual example or rule), and view feature shape plots of selected features, as participants suggested:

> *"If I can click on this (feature attribution) and then I can get this chart (feature shape), I think that would be good. I don't think everyone is going to click it, but I think (if) people want more information, you will click it."* (P20, House)

Many participants tended to check the **local position of their input data point** on the global feature shape diagram.

> *"It's good to see where exactly on a (house price) scale you are."* (P20, House, Trust)

And P30 suggested feature shape could have the assumption that for all the other features that are kept constant, they should be as similar to user's input features as possible.

> *"The AI should assume all the other features are almost the same as mine, considering this hypothesis then this is the (feature shape) curve"*.

### 5.3   Feature Interaction

*5.3.1* **Applicable Explanation Goals**. Since feature interaction just adds one more feature to the feature-outcome plot to show feature-feature interactions, it can be regarded as an expanded version of feature shape, and many of the above findings on feature shape apply to feature interaction as well. Similar to feature shape, feature interaction also supports **counterfactual reasoning** by including two or more features instead of one (as in feature shape).

> *"(feature interaction on age-body weight interaction) If you put yourself in a hypothetical guessing, you're in this age and this is your body weight, and you can already tell the chances (of diabetes) are high."* (P23, Health, Trust)

*5.3.2* **Cons**. *"The graph is **less accessible to understand**"* (P22). In our study, only a few participants could correctly interpret the 2D heatmap of two feature interactions.

*5.3.3* **Design Implications**. Similar to feature shape, participants would like to **choose their interested feature pairs** to check their interactions on the feature interaction diagram. Since the combination of features is large, the XAI system may be able to **suggest interesting feature interactions and prioritize** the feature pairs which have significant interactions.

> "If I click on any two of them (features), show the relationship between them. If I can choose age and blood sugar level, then probably there is some correlation between them. If it is statistically significant, then I would want to know that. If there is no significance between, for instance, age and body weight, then I don't think it should tell me that. If the AI can tell me that this combination really is important for you to look into, then the priority would also make a lot of sense." (P23, Health, Unexpected)

### 5.4  Similar Example

In our study, most participants regarded both similar example and typical example as similar examples. Only a few participants got the idea that with a typical example, *"you're getting the average"* (P20). Thus in this section, we state the themes on similar example as well as the common themes of similar and typical example.

*5.4.1*  **Pros**.  Participants **intuitively understood** the concept of similar example. Similar example uses analogical reasoning to facilitate user's sense-making process.

> "It just intuitively makes sense to me. …similar and typical example are much easier. I don't have to think about them before figuring it out." (P16, Bird, Trust)
>
> "(similar and typical example) It's similar to how humans make decisions, like we compare similar images to the original (input) one." (P02, Bird, Trust)

*5.4.2*  **Applicable Explanation Goals**.  Unlike other explanatory forms that reveal AI's decision-making process (such as rule-based explanations), *"even though these (similar and typical example) aren't much specific about how it's actually doing the (decision) process"* (P16), participants' minds automatically made up such a process by themselves by **comparing instances**. Such comparison mainly allow users to **verify AI's decisions** and to calibrate their trust. The common explanation goals for which similar example were selected are:

**1**) To build trust, especially from a personal and emotional level.

> *This (similar example) made me trust on an emotional level. Because I'm thinking, 'Oh really? I am only 33 years old.' Like I probably not going to get diabetes. But then I'm reading about somebody that does (get diabetes), that sounds a lot like me, it kind of emotionally makes me feel like, 'Oh geez, maybe it is accurate.' So this (performance, output) is like using my brain, and this one (similar example) kind of got me in the gut like, 'Oh, okay. This could actually happen to me. It happened to this person who sounds a lot like me.'"* (P16, Health, Trust).

**2**) To verify the decision quality of AI.

> "It's like a proof for my final decision." (P30)
>
> "Because AI has only 85% accuracy, I want to see similar ones, and what AI thinks they are." (P14, Bird, Trust)
>
> "If it doesn't align (with my prediction), then I want to see some similar houses to remake the judgment."(P04, House, Unexpected).

**3**) To assess the level of disagreement when AI made an unexpected prediction, and to reveal potential flaws of AI.

> *"If my prediction appears in (a list of) similar examples, it allows me to judge whether AI is completely unreliable or just need some improvement."* (P01, Bird, Unexpected)

*5.4.3* **Cons**. Showing examples for comparison may not be applicable when input data is incomprehensible or **difficult to read and compare**.

> *"I think (similar and typical example) it's not important to me. Because I need to read other people's status, read their records."* (P02, Health, Trust)

In addition, participants easily got confused when instances in similar example have **divergent predictions**. This problem might be solved by typical example which is stated in Section 5.5.

> *"(similar example) It's not really telling you if it (the input) is the one (prediction), so it could be this (prediction) or this or this [pointing to different predictions on similar example card]."* (P26, Bird, Trust)
>
> *"This one (similar example) has too many choices (predictions), it's too confusing."* (P05, Bird, Trust)

*5.4.4* **Design Implications**. As mentioned above, participants had to compare the features in similar example by themselves. It is important for the XAI system to support such **side-by-side feature-based comparison** among instances such as input, similar, typical, or counterfactual example, especially when the input data format is difficult to read through.

> *"I don't want to read the text (in similar and typical example), it is better to show those features and examples in a table for me to compare directly, also highlight the important features as an analysis process."* (P29, Health, Trust)
>
> *" Maybe it could help the doctor to pinpoint things that are similar or different between these cases."* (P31, Health, Communication)
>
> *"I would like a comparison. That's my own house (input), which probably will be off the top somewhere. And I'm comparing it with other information (typical example and counterfactual example). So in a column, and I can compare it. For the layout, maybe you can do a product comparison."* (P03, House, Expected)

## 5.5 Typical Example

*5.5.1* **Pros**. One drawback of similar example is that it may make users confused about similar data instances, especially when they have different predictions. Typical example may overcome this problem, since the typical examples for different predictions are more **distinct and separable** than the nearest neighbors of similar examples.

> *"(typical example) You actually made a category of each one. I remember in cognitive psychology, there's a theory. I don't remember the name, but if you clearly separate each category, that helps people to differentiate the different categories, then remember. But for this one (similar example), you have to read every one (instance) of them."* (P04, Bird, Learning)

*5.5.2* **Applicable Explanation Goals**. Since typical example represents the typical case for the outcome, it may help to reveal class-specific characteristics or even potential problems in the AI model or data, for example to **reveal bias**.

> *"If I'm concerned about what group the data is coming from, I would love if the typical case like the average that comes up says like, male, this age, and the factors were quite different from mine, then I

> *kinda go, 'huh?' But if it could give me a typical case that's actually quite similar to me, then I would be less worried about it not performing well with my group."* (P22, Health, Bias)

Unfortunately, most participants did not realize the meaning of typical example and did not make use of such "debugging" property.

*5.5.3* **Design Implications**. In addition to showing typical examples from different predictions (between-class variation), in some cases, it might be beneficial to show **different variations of typical example** for a particular prediction (within-class variation).

> *"It's showing different pictures of the same bird, and the colors even look different. So it's saying maybe, 'Oh, I get it, we have the male and female.' So it's showing different looks that the bird can have."* (P06, Bird, Learning)

Contrary to typical example, some participants expected to see non-typical or edge cases that represent rare but severe consequences, mainly due to safety and bias concerns.

> *"So they (similar and typical example) don't really provide enough information about when the weather is different and when you're driving at night, the results from non-typical conditions."* (P27, Car, Bias)
> *"I still don't know if the dog jumps out of nowhere. so maybe the (similar example) similar traffic conditions can see the extreme cases."* (P03, Car, Safety)

**5.6  Counterfactual Example**

*5.6.1* **Pros and Applicable Explanation Goals**. In our study, counterfactual example was shown as two instances of different predictions, with their feature differences highlighted while keeping other features the same (Figure 1). This format can serve for different explanation goals depending on the task context. In predictive tasks (House and Health), participants regarded counterfactual example as the most direct explanatory form to **suggest ways to improve the predicted outcome**.

> *"For renovations, I think that's (counterfactual example) the only card I would choose. The only one that really tells me that I can do something to increase the price."* (P20, House, Improvement)

Whereas in recognition task (Bird), counterfactual example is suitable to **show the differences** to differentiate two similar predictions.

> *"Counterfactual example let me learn their relationship, highlight the difference between the two (birds). Help me remember the different features."* (P11, Bird, Learning)

*5.6.2* **Cons**. Some participants did not understand the meaning of counterfactual example, and could not capture the nuance between feature attribution and counterfactual example, since they both have features highlighted but for different reasons: feature attribution highlights features that are important for prediction, whereas counterfactual example highlights what features need to change for the potential outcome to happen.

Counterfactual example may have the drawback of making users **confused about similar instances**, especially in recognition tasks.

> *"I think this tool (counterfactual example) will make me remember the wrong thing. I'm already confused. It shows information that is similar."* (P11, Bird)

Thus, it may not be the beginning explanations and may only show up on-demand, for example, for the two explanation goals of improvement and differentiation mentioned above.

*5.6.3* **Design Implications**. The two **contrastive predictions in a counterfactual example can be user-defined or pre-generated** depending on the specific explanation goals. One prediction is usually from user's current instance such as input, and the alternative prediction could be: *"the next possible prediction"* (P18, Bird, Report), users' own prediction when there is a disagreement (unexpected), the prospective prediction to improve the predicted outcome, and the easily confused prediction to differentiate similar instances.

The generating of counterfactual features may also receive user-defined or pre-defined constraints, such as: **1**) constraints on the counterfactual feature type to include **controllable features only** (see Section 6.8 Improvement on controllable features); **2**) generate **personalized counterfactual suggestions** based on features that users look upon: *"the recommendation should be a lot based on what I do"* (P24, Health); and **3**) constraints on the range of specific counterfactual features: *"AI should accept my personalized constraints on budget* (P01, House, Improvement). Given these constraints, the XAI system can also provide multiple improvement suggestions for users to choose from (P01, P11), and may give weights or relative ranking on multiple suggestions.

## 5.7 Decision Rule

Many participants noticed different forms of rule-based explanations (rule, decision tree) provided *"basically the same information"* (P02, Health), *"all show the decision process"* (P10, Bird), and were only different in the text (rule) or graphical (decision tree) representation.

*5.7.1* **Pros**. Several participants regarded rule can *"**explain the logic** behind how the AI makes decisions"* (P27). Particularly, the text description format is *"like human explanation"* (P01, House, Trust), and *"simple enough and understandable"* (P11).

*5.7.2* **Applicable Explanation Goals**. The above pros make rule suitable for verbal (Section 6.9 Communication) and written **communications** (Section 6.10 Report). Text format may also help to dispel confusion, since some participant regarded texts as being more precise than images, thus **facilitate learning**.

> *"In this case (Bird, Unexpected), I don't want to see the highlights (feature attribution). I want it to see points, the specific parts and give me some explanation. If I'm trying to prove myself wrong, or if I want to see how AI system can prove me wrong, I want to see more precise text, and precisely point out the important information."* (P04)
> *"The written helps because it's more exact, whereas the pictures, ...the blue in the picture might not be the blue that was in the written."* (P05, Bird, Learning)
> *"(rule) It's listing out something that a person might miss in the picture."* (P18, Bird)

However, when the input is image data, some participants also mentioned providing text explanations only was not enough.

> *"(Rule) It doesn't really show you the bird that you were looking at. Lots of birds have small thin bills short tails...if I can't see a picture of it, then it's not as helpful."* (P06, Bird, Trust)

And many participants suggested *"ideally you'd want both written and pictures"* (P05) to complement each other.

EUCA: the End-User-Centered XAI Framework                                                                29*5.7.3* **Cons**. Rule is **very sensitive to the degree of complexity** in text descriptions, as an increase in rule length or number of features will dramatically reduce its simplicity and the above advantages [74]. However, if the rule clauses are short, the explanation may not be precise and satisfying as well, as P06 pointed out,

> *"It (rule) is just too broad, it could apply to so many other birds."*

Another concern is that since participants lack technical knowledge, some of them misinterpreted rule as instructions human fed to the AI.

> *"(Rule) it is giving very clear instructions to the AI, like written text instructions, these are already fed into the system."* (P09, Car, Safety)

*5.7.4* **Design Implications**. To reduce the cognitive load of complex rules, a few participants suggested **trimming the rules** by presenting shallow levels only, or *"just show rule related to my own house features"* (P30), and users may query details on demand.

To carefully balance between explanation completeness and usability, if the full rules are shown, it is beneficial to **highlight *local* rule clauses** describing the current instance on top of the *global* rule explanation.

## 5.8 Decision Tree

*5.8.1* **Pros**. Similar to rule, participants regarded decision tree as *"the most logical one"* (P20) that *"**tells you the decision-making process**"* (P04):

> *"(Decision tree) shows the process of thinking with AI, what it's going to do with the information."* (P10)
> *"How the algorithm is working, what the machine is thinking about when it's coming up with the prediction."* (P16)

*5.8.2* **Applicable Explanation Goals**. Participants mentioned an advantage of decision tree is to **differentiate similar instances**, possibly due to its unique tree layout:

> *"It explained very well what's the difference between them (the two confusing instances)."* (P04, Bird, Report)
> *"It would show you how to pick up the different types of variants."* (P10, Bird, Report)
> *"I think this (decision tree) is the graphic comparison, like this beak might be sharper or smaller than this one, all those comparisons help"* (P09, Bird, Unexpected).

Such advantage also supports **counterfactual reasoning** by checking alternative feature values on the adjacent branches.

> *"(Decision tree) can see how to improve. It has a comparison with different outputs."* (P29, Health, Trust)
> *"Where does my house stands, if I'd be here, then I maybe try to change some of my features, to see how do these features affect my house price, or other houses compared to my own house."* (P30, House)

*5.8.3* **Cons**. Several participants brought up its **weakness in communication and interpretation**.

> *"(Decision tree) is not natural language, it is more difficult to explain to my family."* (P01, House, Communication)
> *"This is more like a logical thing for me to see. But I wouldn't use this as an explanation to family, because that's just weird. I don't want to rack their brains too much."* (P20, House, Communication)

Indeed, in our user study, even with a two-feature two-layer decision tree, a number of participants commented:

> *"It's confusing."* (P05, Bird, Learning)
> *"It got too much information."* (P16, Bird, Unexpected)
> *"I don't really understand this one. I think it's a little bit complicated."* (P08, Bird, Learning)

Since it is less interpretable than other forms, some participants suggested to show it on-demand.

> *"I don't think these two (decision tree, decision flow chart) are necessary to show in the first UI. Maybe these two can be hidden in an icon that says 'process'. Because it (decision tree) is more like a program in process."* (P04, Bird, Trust)

Besides the tree structure, we used another flow chart visual representation (decision flow chart) in the study. In tasks that the input data were images (Bird and Car task), quite a few participants found neither the tree nor the flow chart structure helpful, and they only focused on the saliency features or objects in the flow chart.

> *"I don't think it (the flow chart structure) matters, just the head and the belly (the highlighted region shown in the flow chart) matter."* (P14, Bird)

*5.8.4* **Design Implications**. Similar to the suggestions in rule, to reduce its complexity, one participant suggested **trimming the tree** and just showing the main branches, hiding the deeper branch and only **showing details on-demand**.

> *"You could use this one (the two-feature decision tree) as a beginning, based on this, and you click (one branch) to another in-depth version of the price calculation. Because this (price prediction) range is still very far wide, and the features given is not enough, so if you want to (check details) maybe click and (it will) add more features to it (that branch), then get a narrow range (of prediction)."* (P28, House)

Although rule-based explanations are *global* explanations (on model's overall behavior), many participants tended to focus on the branch pathway where their own input resides. By doing so, they were seeking *local* explanation (of the current input) on top of the *global* explanation. This suggests a tree-based explanation may only need to show branches containing interested instances, or **highlight branches for user's interested instances**. Users did so for the explanation goals to verify AI's decisions, and to compare with other counterfactual instances.

> *"I know there're factors that could be other houses that lead to different prices, but I still see it as, 'okay, I plug in my own numbers here and what's my price?' So it's still specific to me.'* (P20, House, Trust) –
> Displays local as well as global explanations
> *"The only thing we need is to indicate my own position on this (decision tree) branch. ….Then I can chase the features of my house."* (P30, House, Unexpected) – Suggests to highlight the pathway for user's interested instance

## 5.9 Input

It serves as necessary background information, and participants regarded input as a *"profile"* (P24) that *"stating the facts"* (P20). It allows participants to **understand what information AI's decision is based on**, and can help "debug" to see *"if AI is missing the most important feature"* in input (P22, Health, Bias), and *"whether or not the input is enough for it (AI) to make that decision"* (P16, Health, Trust).

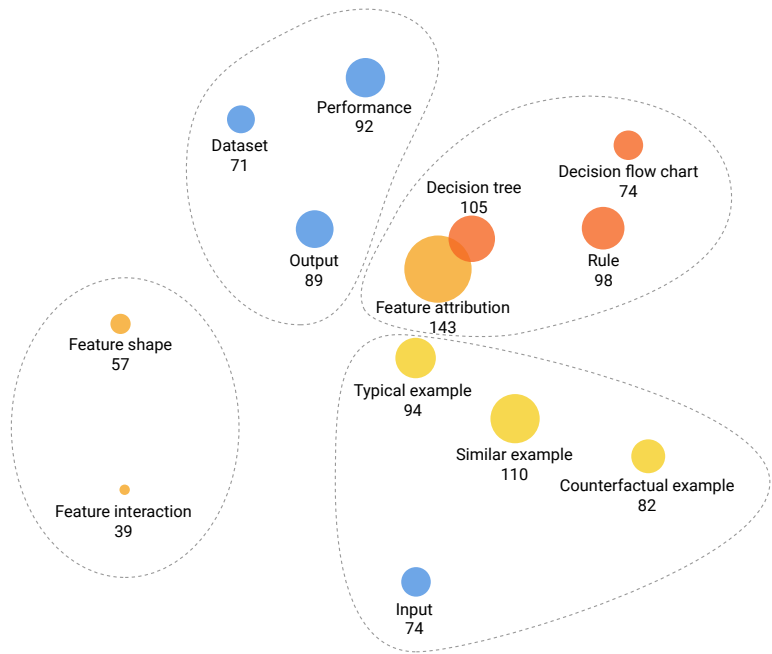

Fig. 8. **Visualizing the similarities of the 12 explanatory forms in EUCA**. The explanatory forms that are close to each other indicate they are more likely to be selected together to construct an explanation. The total number of times an explanatory form was selected is indicated as the number below its name, also proportional to its dot size. The dot color indicates its category in the EUCA framework: feature, example, rule, and supplementary information. The categorical clustering roughly corresponds to the automatic k-means clustering (circled together with dotted line), which is based on the pairwise similarity matrix measured as the co-occurrence of a pair of cards in card selections from the user study.

When checking input, participants tended to intuitively *"look for certain features"* (P14) to judge by themselves. In the card selection & sorting, some participants used input as an anchor, put it side-by-side with example-bases explanatory forms (similar, typical, and counterfactual example) for comparison. Quantitative results led to the same findings, as input was clustered together with other example-based explanatory forms (Fig. 8).

### 5.10 Output

In our study, the output card contains prediction information of a point prediction, a prediction range, and their corresponding uncertainty level (for regression tasks); Or top three predictions and their likelihood (for classification tasks) (Fig 1). For the output information presentation, some participants preferred to check the point prediction at the beginning, and check the detailed prediction range and uncertainty level on-demand or leave them at the end, since they *"need a longer time to understand what these numbers mean"* (P02).

Participants had divergent preferences and understandings on the prediction presentation forms. Compared to a point prediction (e.g.: the prediction on house price is 650k), some preferred to see a **prediction range** in regression tasks (e.g.: house price is 638-662k), or top predictions list in classification tasks, because such prediction range *"give choices"* (P05, Bird classification task, Differentiation), *"acknowledges a possibility"* (P18, Bird, Unexpected), rank the decision priorities (P03, Car classification task, Safety), help them *"(the range) to see how different between my and AI prediction"* (P01, House regression task, Unexpected), and provide rooms for adjustment and negotiation:

> *"If I want to sell it higher, and I'll put 662k (the upper bound). Or if I wanted to sell it fast, then I'll put 638k (the lower bound). There's always a range, it's not necessarily just one price. And people will always bargain too."* (P20, House, Communication)

And sometimes they *"don't even need to know the (prediction) number exactly. This (range) tells me that (my diabetes risk) it's high. I have to do something. So that's what I want to know"* (P17, Health regression task, Trust), and the range gives a higher certainty than a single point prediction which enhanced participants' trust.

In contrast, some other participants were more acceptable to a narrower range or a **point prediction**, because they saw a wider range of prediction had its drawbacks: *"(the prediction range) shows too much fluctuation"* (P07, House, Trust); And seeing the full predictions list (some with lower prediction likelihoods) may make them confused and discredit AI's decisions. Thus a narrower range may give them more confidence about AI's prediction.

> *"Seeing that the range is pretty small makes me a lot more confident that they've got enough data to actually be drawing conclusions."* (P16, Health, regression task, Trust)

For the prediction likelihood/uncertainty/confidence[1], some participants **had a hard time understanding the meaning of uncertainty** and required researcher's further explanations. A high certainty *"reassure AI's performance"* (P22), *"help a lot of persuading yourself into believing in AI"* (P10), which is consistent with the recent quantitative finding on certainty level and trust calibration [104]. Especially when AI and users disagree with each other (unexpected), participants may abandon their own judgment due to AI's high certainty.

> *"If it had a high certainty, then I would want to know why my estimation is wrong."* (P10, House, Unexpected)

### 5.11 Performance

After checking the performance information, most participants realized the probabilistic nature of AI decisions: *"AI is not perfect"* (P20), *"they (AI) make errors sometimes"* (P05). If the performance is within their acceptable range, participants would accept the "imperfect AI". And seeing the performance level helped users to **set a proper expectation for AI's performance**.

> *"I get it's downside. Performance warns me to, 'Hey, you know, it's not really accurate. There's some room for error.' "* (P24)

And sometimes participants may calibrate their trust according to the error rate (in classification tasks) or error margin (in regression tasks).

> *"If there is a really big margin (of error), then it would probably demean the trust."* (P23)

Almost all participants understood the meaning of accuracy (error rate) in classification tasks, whereas many participants had a difficult time understanding the margin of error in regression tasks.

> *"Performance is really in detail. I mean not everyone is familiar with statistics, like mean error."* (P30, House, regression task)

---

[1]Although the AI community has distinct methods to compute output likelihood and uncertainty level, in our study we used likelihood, confidence and uncertainty interchangeably to avoid participants' confusion.

Unlike the uncertainty level in output (Section 5.10) which is case-specific decision quality information, a few participants noticed performance is model-wide information, and just provides *"general information showing the trust level of the system"* (P04) is *"too general, I would want to know specifically why (the speed) it's going down in this particular case of driving"* (P05, Car, Unexpected). Thus, they suggested there was no need to show it every time, *"you should know before you use AI"* (P11).

However, in some particular explanation goals such as to detect bias (Section 6.3), participants may require to check the fine-grained performance analysis on interested outcome.

> *"It (fine-grained performance on road/weather conditions) explains how often I should be confident in rainy days."* (P19, Car, Safety)

### 5.12 Dataset

In our study, the dataset card contains training dataset distribution of the prediction outcomes. Even after researchers' explanation, some participants did not well understand or **misunderstood** the information on this card (for example, some misinterpreted the distribution graph as feature shape), indicating it requires a higher level of AI/math/visualization literacy [21, 37, 66]. For those who comprehended the dataset information, some participants tended to link the dataset size with model accuracy and trust.

> *"The higher the (training data distribution) curve goes, then I would be more confident that they have a big pool of data to pull from."* (P31, Health, Unexpected)

Some participants intuitively wanted to check their own data point within the training data distribution, and use it as a dashboard to **navigate, identify, and filter interested instances** (such as similar, typical, and counterfactual examples), to compare what are the same and different features between their input and the interested instances.

> *"I want to see which region I fall in the population, and compare with people around to see why my (diabetes) risk is only 10% with a family history."* (P01, Health, Unexpected)

Nevertheless, in practice there may be some restrictions on reviewing the detailed dataset information due to data proprietary and privacy, as brought out by P19:

> *"I want to know the number of data and the details of it to verify. But I don't know if that's going to be able to be viewed. That's probably secret, right?"* (P19, House, Expected)

Table 4. Key findings on RQ2: end-users' requirements for explainability under various explanation goals.

| Explanation Goal | End-Users' Requirements for Explainability |
| --- | --- |
| **Calibrate Trust** | Users require stated and observed performance; Request features which the decisions are based on; Request information on the capability and credit of AI. |
| **Ensure Safety** | Performance in various testing cases to show AI's robustness on safety. |
| **Detect Bias** | Require AI to maintain the same capability and perform well for minority subgroups. |
| **Unexpected Prediction: Disagreement with AI** | Users may lose trust; Need explanations for verification; User may check input for debugging. |
| **Expected Prediction** | Users may not need explanations; Or need explanations to boost decision confidence, and to improve the current outcome. |
| **Differentiate Similar Instances** | Be able to discern similar instances and pinpoint the feature differences; Indicate decision certainty level. |
| **Learn from AI** | Built upon reliability and trust of AI; Require a wide range of explanation to support user's own learning. |
| **Improve the Predicted Outcome** | Require solid evidence on improvement suggestions; Seek changeable features. |
| **Communicate with Stakeholders** | Explanations depend on different stakeholders and goals of communication. |
| **Generate Reports** | Explanation content depends on specific explanation goals and readers of the report. |
| **Multi-Objectives Trade-Off** | Allow users to take over; Users use explanations to defend for or against certain objectives. |

## 6 RQ2: END-USERS' GENERAL REQUIREMENTS FOR VARIOUS EXPLANATION GOALS

In the previous section, we answer the main research question on end-users' perceptions of the 12 explanatory forms. These findings were elicited within the context of different explanation goals. Next, we report results on end-users' general requirements for these explanation goals (mainly from Round 1, and partially from Round 2 of the user study). The key findings are summarized in Table 4. The findings serve as design reference to help designers understand end-user's potential requirements, so that they could prepare more detailed questions and have a profound conversation with their stakeholders.

Although the range of explanation goal is wider for end-users than for technical users, two main themes of explanation goals emerged in the interview. Quantitative clustering analysis confirmed similar trends as visualized in Supplementary Material S2 Fig. 6.

The first and fundamental driving force of checking explanations is to **verify AI's prediction**, and to gain trust and understanding of the AI system.

> *"Like boyfriend and girlfriend, I want to know what my boyfriend is thinking. Similarly, I want to know what the car's thinking before I'm with the car."* (P32, Car task, Safety) – To gain understanding on AI
>
> *"I will also want to know how the software can predict the 80% chance that I'm going to have diabetes. And also, how did they come up with that numbers? Just giving me a number without justification or some verifiable reasons, it's just unlikely I would accept it because it may not be true."* (P25, Health task, Trust)

The following explanation goals are more related to this motivation: calibrating trust (Section 6.1), ensuring safety (6.2), detecting bias (6.3), and resolving disagreement (6.4).

The second motivation to check explanations is for **personal improvement**, i.e.: to improve users' own welfare, such as to enhance personal problem solving skills and learning, or to control and improve the predicted outcome. This is built upon the trust and result verification from users' prior experience with AI, as one participant stated:

> *"(When trust has been established,) what has to be done in the next phase of the (AI) software, is how the software is being helpful to me. ...If I know the result, I don't think I would want to dig in to see why it is, but I would want to see how I can reduce the chances of diabetes."* (P23, Health)

The following explanation goals are more related to this motivation: to seek suggestions to improve the outcome (6.8), to learn and discover new knowledge (6.7), to differentiate similar instances (6.6), to facilitate verbal (Communication, 6.9) and written communication (Report, 6.10), and to balance among multiple objectives (6.11).

## 6.1 Calibrating Trust

The process of calibrating trust involves multiple factors and their complicated interactions. We summarize the following key emerged themes participants requested to calibrate their trust toward AI.

(1) **Performance** Trust towards AI is fundamental to incorporate AI's opinion into the critical decision-making process [101], and many other explanation goals are built on trust.

   Since end-users usually do not have complete computational and domain knowledge to judge AI's decision process, model performance becomes an important surrogate to establish trust.

   > *"Even if AI tells me how it reaches its decision, I cannot judge whether it's correct since it is a medical analysis and requires professional medical knowledge. I just know the accuracy and that'll be fine."*
   > (P01, Health)

   Prior work identified two types of performance: stated and observed performance [99], and both were mentioned in the interview. *Stated performance* or accuracy is performance metrics tested on previous hold-out test data, and it was mentioned by most participants as a requirement to build trust towards AI.

   > *"I understand maybe AI is learning from past examples, and it may be finding patterns in the data that might not be easy to explain. So I'm less concerned about how it's getting there. I think I do have a trust that is doing it right, as long as there's something you can test after how accurate it's been."*
   > (P16, Health)

Compared to assessing the performance metrics, some participants tended to test AI by themselves and get hands-on *observed performance* to be convinced. This requires users to have a referred ground truth from their own judgment or reliable external sources.

> *"(My own test driving experience) is way more useful than watching a (test driving) video, because you shouldn't trust everything. The video might be just made to wait for you to buy the car. So talking from a customer perspective, I would like to try it myself, because I also sell things. So I would always like to try it myself instead of watching a video."* (P21, Car)

(2) **Feature** The important features that AI was based on are the next frequently mentioned information.

> *"I would like to know the list of criteria that the AI chose the price based on, and which one weighs more."* (P30, House)

(3) **The ability to discriminate similar instances** This information was requested by several participants to demonstrate AI's capability.

> *"(Decision tree) It's showing me that it's picking from a few similar ones, not just like a random ray of blue, purple, green birds. It's not random, it's a calculated response. More of that would help me trust AI."* (P06, Bird)

> *"Typical example seems to be pretty good at picking up on differences. Similar example I can see that it's got a good variety of similar birds. So I found these ones make me trust it more.* (P16, Bird)

(4) **Dataset** The dataset size that AI was trained on is another surrogate mentioned by some participants to enhance trust.

> *"To me what artificial intelligence does is just collecting a lot of data, and tries to make sense for behavioral patterns. So I would actually trust it, because I think it's just based on data, it is a more accurate measurement of what market rate is for house prices."* (P03, House)

> *"If I know that the AI comes from a large database, it seems like the database is actually the experience that AI has. So the larger it (dataset) gets, the more experienced AI would be, so I can trust it more."* (P30, House)

(5) **External information** This is another surrogate mentioned by participants to judge if AI is trustworthy. The external information could include:

   (a) Peer reviews, endorsement, and AI company's credit.

   > *"Since I'm not really a tech person, so I'm not sure how I look at it in a technical way. So that's why I just really depend on the company's reputation, and also how people feel about the website."* (P28, House)

   (b) Authority approval and liability.

   > *"I trust more if the government themselves kind of stands behind it, getting some sort of government approval helps it a little bit more. So if there's some health authority like Health Canada or FDA support gives it more legitimacy."* (P24, Health)

   > *"For me personally, I would prefer if an actual person is there in the end, at least in the beginning stage. So if somebody is there to just say, 'hi, I'm so and so', and then AI takes control. Then we still know that there is somebody who's liable in the end for whatever happens."* (P23, Health)

### 6.2 Ensuring Safety

To ensure the safety and reliability of the AI system in critical tasks (the autonomous driving vehicle task in our study), participants frequently mentioned **checking AI's performances in test cases**, expecting the testing to **cover a variety of scenarios** to show the robustness of safety. Although it is impossible to enumerate a complete list of potential failure cases in testing, extreme cases or potential accidents were the main concerns and focus of end-users.

> *"Potential crashes or just like someone speeding or a pedestrian jumping out of nowhere."* (P19)
>
> *"There is likely to be someone running around, so it needs to show me the extreme cases. ...I need to see something like FMEA, failure modes and effects analysis, just to be like, 'okay this is how it works.' because I know nothing is foolproof. There are always to be something, but to what extent."* (P03)

Similar to the explanation goal to [calibrate trust](), alongside the above *stated performance*, a few participants required *observed performance* to emotionally accept AI as an emerging technology.

> *"Definitely I would want to be in one car. I think information is not helpful, it's not an intellectual factual thing, it's emotionally not acceptable. It (AI) is new and I have to learn to trust it."* (P17)

### 6.3 Detecting Bias

Participants were concerned about population bias [70], or distribution shift where AI models are applied to a different population other than the training dataset. Such concern is more prominent when a prediction is made on users' own personal data, and when users are in minority subgroups. Participants wanted to compare and see **if their own subgroup is included in the training data**.

> *"I know I'm in a class, they talked about how a lot of studies haven't been done specifically on women, even though they (diabetes) affect men and women differently. That is probably something I would want to know about, like if it gave me this result and then it had a little note that explained the research was done more on that demographic, so it may be more true for that demographic, but they're just trying to, what's the word, extrapolate to this group where I sit."* (P22, female, Health)

Unlike the common bias and fairness problem in AI where the protected features should not affect the prediction, in our Health task on diabetes prediction, the protected features (age, gender, ethnicity) do lead to a difference in diabetes outcomes (referred as explainable discrimination in [70]). Participants who were aware of this point **required AI to account for such differences among subgroups**.

> *"I know some ethnic groups just by genetic makeup could be more predisposed to diabetes. In order for it (AI) to arrive at this decision, I would think that it has maybe like a sample size of different people with different ethnicities to try to figure out. I would think there'll be years and years of research has already been done of the different groups, different ages that would then be factored in by AI. If I can see it (AI) is using that information, I'll be a lot more comfortable to actually using the AI's recommendation."* (P17, Health)

In cases where the AI task is not related to personal information (in our study the self-driving car task), participants required AI to be able to detect objects and perform equally in all potential biased conditions.

> *"Now we are operating in night time, or different weather, but they (the self-driving cars) still have to be able to see the signs and identify the objects."* (P13, Car)

### 6.4 Unexpected Prediction: When Users Disagree with AI

When AI's predictions did not align with participants' own expectations, most participants would *"question AI"* (P16, P20) and *"the contradiction may let me confuse"* (P02). Some may lose trust in AI, thus would not go further to check its explanations, if they were confident about their own judgment. Some would check *"a trusted second opinion"* (P06, P10), or refer to human experts (P12).

But for the majority of participants, **explanations were needed** *"to know why"* (P01) and to resolve conflicts.

> *"I'm feeling conflicted because it's giving me two different information, my own personal belief and AI. So in order to convince me that AI does know what it's talking about, you need to go through the mental validation step [pointing to the ranked explanatory cards]. So by the time I go through this (explanatory cards) and I come out of it, I am extremely convinced."* (P24, Health)

Explanations help users to identify AI's flaws and reject AI, or to be convinced and adjust user's own judgment by checking detailed differences on rationale, although *"it might be harder to persuade me"* (P31). Specifically, participants *"try to understand what makes a difference (between AI and my prediction)"* (P03), which is similar to the explanation goal to differentiate (Section 6.6). To show why the predictions are different, many participants **required a list of key features**.

> *"Because AI cannot think like a human, so the reason that I ask for the criteria list is trying to think how similar to me is AI's thinking. So maybe AI is thinking better, or is seeing a wider range, so it's checking things that I've never thought about."* (P03, House)

In the case that AI made errors, seeing what AI is based on can facilitate user's "debugging" process. Although end-users cannot debug the algorithmic part, they may be able to debug the input to see if AI *"have the complete information"* (P03) as users have. Furthermore, if some key input information is lacking in AI's decision, the system needs to allow users to provide feedback by inputting more information (P03, P24), or *"correct the error"* (P16) for AI.

### 6.5 Expected Prediction: When Users Agree with AI

In contrast, when the prediction matched participants' expectations, participants *"will trust the AI more"* (P10), and the motivation to check explanations was *"not as strong as the previous one (unexpected)"* (P02). Some participants stopped at the prediction, willing to accept the "black-box" AI and may *"not even waste my time (checking explanations)"* (P20).

A few participants still wanted to check further explanations for the following motivations:

(1) To **boost user's confidence**.
> *"Even in this (expected) scenario, it would be nice to have some bullet points, like the reasons behind it the estimation being accurate, because if someone says that you're charging me way too much, I can have point by point reasons explaining to you why this house worth this price, it actually kind of as a confidence boost to think you are not overcharging or undercharging."* (P03, House)

(2) To **improve the predicted outcome** (see Section 6.8 for more findings).
> *"If diabetes already runs in my family, (and AI predicts my risk of diabetes is 80%), it would probably make me more confident about the software. So I might want to ask for more information about which aspects of my health records were the most important for making this decision? Coz then maybe that can help me with my future activities and changing things in the future."* (P31, Health)

### 6.6 Differentiating Similar Instances

To facilitate end-users' explanation goal to differentiate similar instances, an AI system is required to first have the ability to discern similar instances.

> *"Depends on how good it is...So I think you would have to improve how AI picks up the birds, like maybe these are the same color birds, but maybe they have slightly different characteristics. So if AI can pick that up, then I think it would be better."* (P10)

And in case of doubtful prediction, participants expected AI to indicate how certain it is to the prediction.

> *"I would expect AI if it doesn't know, it would give choices. So it would say 100% or 99% that's an indigo bunting, and 89% it thinks it's a finch."* (P05)

Based on that, AI needs to be able to *"**pinpoint unique features** that made them really different from each other"* (P06). In addition, the interface may also need to support users' own comparison.

> *"AI can tell you what the differences are. I guess it could be some list of the beak is longer for this and that. But I think visually bringing the differences up side by side, and then I can directly compare what the differences are."* (P16)

### 6.7 Learning from AI

Using AI for user's personal learning, improving problem solving skills, and knowledge discovery, *"depends on how reliable it (AI) really is"* (P10). And participants expected AI to *"receive human feedback to correct its error and improve itself"* (P01).

To facilitate learning and knowledge discovery, *"just looking at (input) pictures and (output) names isn't enough"* (P10), and participants **expected a wide range of explanations** depending on the particular learning goal, such as *"more details to systematically learn, go over that same bird, ...a mind map to build a category of birds by one feature"* (P02), *"the specific characteristic about this bird, and how can I differentiate this bird from other birds"* (P04). Other learning features mentioned by participants include: referring to external *"respectable source"* (P18), supporting personalized learning for unfamiliar terms (P04), and *"collecting information about how well I'm doing on it, like if I guess wrong, does it record that? to see if I'm progressing"* (P16).

### 6.8 Improving the Predicted Outcome

Participants intuitively sought explanations to improve the predicted outcome, when predictions are related to personal data (in our study, the House and Health tasks). However, they **tended to unwarrantedly assume the explanations were causal** (causal illusion [69], i.e., believe there is a causal connection between the breakdown factors and the outcome), even though the cause-effect relationship has not been confirmed, and AI largely relies on correlation for prediction [76]. Only a few participants required more solid evidence to support the explanations on improving the predicted outcome, especially when the action was related to critical consequences (personal health outcomes).

> *"I presume the recommendation (on improvement actions from AI) is also has been backed up by Health Canada, because I think I would tend to follow the recommendations if I know there's definitely medical support behind it."* (P24)

> *"I would definitely want to know like what can I do to mitigate those risk factors or to address those things so that I can decrease the risk. I would really like to know if it had an explanation of how reliable each source was. Coz I know some studies, they might seem like a correlation, but it doesn't mean it's a direct cause. So I would really love it if it could potentially explain how powerful those studies are suggesting."* (P22)

Regarding the specific requirements on the explanations for improvement, participants were looking for **controllable features** and ignoring the features that cannot be changed.

> *"I can not change my age, but I'm able to reduce my weights."* (P02)

Knowing the controllable features has a positive psychological effect to give users a sense of control, and vice versa.

> *"If I'm afraid of getting diabetes, and assume that I'm going to sentence, it feels like there's nothing I can do about it. But when I see this one (feature attribution), I think, 'oh geez, maybe there are other factors here that I can do something about.' So this may make me more positive about doing something about my condition."* (P16)
>
> *"I know it (feature interaction) is comparing my house area and my number of rooms with other houses. I can understand 'okay if I increase my room number, the price will be increased that much.' But the problem is I cannot change any of them (the house features). It just gives me the feeling of disappointment."* (P30)

To counterpoise the unchangeable features, users may intuitively apply counterfactual reasoning to compare different feature adjustment settings.

> *"If I make any change in my house appliance and renew, then I can still reach the same price as if my house was bigger"* (P30).

### 6.9 Communicating with Stakeholders

To communicate with other stakeholders, some participants chose to communicate verbally about their opinions without mentioning AI. Others preferred to present stakeholders with more evidence by bringing AI's additional information explicitly to the discussion. For the latter case, the **other stakeholders need to establish basic understanding and trust towards AI** before discussing AI's explanation.

> *"I'd sit down and get my family together and explain about the artificial intelligence thing."* (P12, House)
>
> *"I would try to get some evidence from it (AI) that I could take to the doctor to get them to buy into it."* (P16, Health)

To do so, most participants chose to present AI's performance information to build trust.

> *"As long as the backstage is accurate and then I can just provide accuracy to my wife and she'll be able to get that. Trustworthy is the most fundamental."* (P28, House)

Different audiences and communication goals may require distinct explanations, as described by P32:

> *"I'm pretty sure my husband or my mother has a different way to decide or they want to know different things."*

In addition, in the Health task, we asked participants to communicate with family members or doctors about their diabetes predictions. Participants' requested explanation covered a wide range of contents, and we did not identify any significant differences in the communicating contents between the two audiences.

A formal summary or report from AI may facilitate the communication with other stakeholders, as requested by many participants.

> *"A written report from AI that I would be able to reference to, in order to talk to my family about that. It would feel a little bit more official rather than just, 'oh, this is what somebody said', there's no real evidence, whereas this sort of creates that paper trail."* (P31, Health)

### 6.10 Generating Reports

The content of reports may largely depend on the specific explanation goal and readers of the report. In our study, participants frequently mentioned the report should include *"key identifying features"*, *"list of distinguishing characteristics or what makes it unique"* (P09), or **"a summary of factors** *that were part of the input led to the diabetic prediction"* (P31). Users also mentioned including supporting information to back up the decisions, such as the training dataset size of the predicted class, and the decision certainty level (P01).

### 6.11 Multiple Objectives Trade-Off

Usually it is the human user rather than AI to trade off among multiple objectives in AI-assisted decision-making tasks. Thus, when multiple objectives conflict (in our study, they are the scenarios when car drives autonomously and passenger gets a car sick; AI predicts diabetes and uses it to determine insurance premium), AI is required to allow users to take over or to receive users' inputs.

> *"It's the most important thing I would want to do is to allow me to stop, or asking to slow down if I'm feeling sick."* (P03, Car)

Explanations are required if the multiple objectives conflict and need to trade off. And users could use such explanations to defend for or against certain objectives.

> *"I think it's like a defensive thing, like if I'm expecting that they're going to cause an increase in my payments or whatever they're going to deny me (health insurance) coverage, I would be trying to find out what it's based on for the opposite reason maybe to discredit it."* (P16, Health)

## 7 DISCUSSIONS

In the previous two sections, we presented the user study findings on end-users' perceptions of the twelve explanatory forms in the EUCA framework (Section 5), and end-users' requirements for various explanation goals (Section 6). Next, we state how could HCI/AI practitioners and researchers utilize the findings and the EUCA framework, and compare the findings with prior literature.

### 7.1 Utilities of the EUCA Framework

*7.1.1 Explanatory Forms as Building Blocks.* In the user study, participants selected and sorted the explanatory form cards, and combined them to construct a low-fidelity prototype. Participants found that the resulting prototype fulfilled the majority of explanation goals (231 out of 279 responses, 83%). It shows the explanatory forms in EUCA may serve as building blocks, and combining them can complete an explanation. The finding resonates with previous user studies that an XAI system should support "integrating multiple explanations", as "users employed a diverse range of explanations to reason variedly" [89]. The combination helps to overcome the weakness of an individual explanatory form, and may make the explanation more robust, complete, and versatile. With multiple explanatory forms that complement each other, users will depict a holistic picture about AI's decision process, and may mitigate confirmation bias, attribution bias, and anchoring bias [61].

To determine the ideal combination of explanatory forms, designers will need to work closely with end-users or other stakeholders, as detailed in the next section 7.1.2. The combination can be *fixed* or *dynamic*: Different explanatory forms can be combined *statically* as different modules in the UI, or *interactively* combined and incorporated in the UX to show detailed explanatory forms on-demand [29, 85, 86]. The contents of combination can be *fixed*, or *dynamically* generated according to user's question, i.e., the XAI system learns to use different explanatory forms as vocabularies to respond to user's follow up questions, so that to construct an interactive explanatory conversation [91] with end-users.

*7.1.2* **Prototyping with EUCA**. Our user study illustrated the proposed prototyping and co-design process with end-users. To design prototypes for a particular XAI application, we summarize the prototyping process from our user study to determine the most feasible combination of explanatory forms, and suggest the following co-design and prototyping workflow (Fig 3).

(1) **Create prototyping cards from explanatory forms**

The designer starts by **manually extracting several interpretable features** given the AI task and input/output data type. For example, for tabular data, the features could be the column names that describing the input instance, such as house size, age, and location. For image data, the features could be saliency image part or object for recognition, such as cars, traffic signs, or pathological appearance of a disease on chest X-ray. As quick prototyping, the feature content may not necessarily reflect the real content generated by XAI algorithms. They serve as content placeholders to be filled in the prototyping card templates.

Then the designer can use the prototyping card template provided by EUCA, and **fill in the template with the above extracted features**. The design templates visualize the explanatory forms as UI elements. Designers can also create their own templates from scratch by referring to the EUCA design examples. The EUCA framework website[2] encourage the community to share and reuse of design patterns for similar XAI applications.

To encourage brainstorming and divergent thinking during the conversation with end-users, the designer may **prepare multiple variations** for some particular explanatory forms, by varying its visual representation (e.g.: graphics or text) and UI layout, alternating contents from brief to details, and providing different options, such as whether to use pre-defined or user-defined contrastive outcome on counterfactual example, whether to give users the option to set a threshold level for feature attribution, or refer to the UI/UX design implications part in the user study findings (Section 5). Each explanatory form and its variations are presented on individual prototyping cards.

While designing UI/UX variations of the prototyping cards, designers may also consider **applying the general human-AI interaction guidelines** (see [13] for details). We selected the following design guidelines that are more relevant to XAI system: "remember recent interactions", "support efficient invocation, dismissal and correction", "learn from user behavior", and "encourage granular feedback".

---
[2]http://weinajin.github.io/end-user-xai

(2) **Co-design and iterate low-fidelity prototype with end-users**

With the prepared prototyping cards, the designers then can meet and discuss with the target end-users and/or other stakeholders of the XAI system, and apply user-centered methods (informally or formally), such as: interview, focus group, and card sorting. The communication aims to use the created cards as a prototyping tool to understand users' needs and requirements for their explanation goals, and incorporate end-users' perspectives in the prototype co-design, iteration, and evaluation process.

To quickly create a low-fidelity prototype from the prototyping cards (prepared in Step 1), the **end-users can select, rank, combine, modify the prototyping cards**, or sketch new ones. In this process, designers may ask users why they selected or did not select a card, and their rationals for making such a combination, whether the combination could fulfill their requirements, and what is lacking in the current prototype. Users can also comment on and revise each variation of the same explanatory form. With the tangible prototyping cards, designers can know in-details about users specific requirements on the XAI system.

Beside understanding users' needs, the prototyping cards can facilitate the UI/UX design. Users can easily manipulate the cards to examine different UI design possibilities, or arrange the layouts as mock-ups on different devices (computer, tablet or mobile phone). For the UX design, users may choose to hide some cards and only show them on-demand, or dynamically present different explanatory information for current context.

After the initial communication with users, designers need to synthesize users comments and decide one or several prototype designs (such as using majority voting). Then based on the prototyping card ranking and combination, the designer may create low-fidelity prototypes, and continue to seek user and/or other stakeholders' feedback and iterate the prototypes.

During the above process, the designer may refer to the user study findings to be informed about the properties of the explanation forms (pros, cons, applicable explanation goals, and design implications in Section 5), and to understand end-users' diverse explanation goals (to calibrate trust, detect bias, resolve disagreement with AI, etc. in Section 6).

(3) **Implement a functional prototype**

After the above co-design process and several rounds of iteration, the low-fidelity prototype will be ready to implement. Based on the selected explanatory form in the prototype, the development team can use Table 3 to identify their corresponding XAI algorithms, and refer to open-source toolkits (such as [2, 4, 5, 8, 9, 16]) to implement high-fidelity functional prototypes.

*7.1.3 Insights for Novel XAI Algorithms/Interfaces.* Our findings provide design implications and insights from end-users' perspectives. It would motivate HCI and AI researchers to develop novel interfaces/algorithms for end-user-oriented XAI. We give some examples for inspiration:

- The design implication in similar example (Section 5.4) indicates users need the system to pinpoint corresponding features among similar examples for easy comparison. This requirement can be regarded as a combination of two explanatory forms: similar example and feature attribution. Such insight may inspire UX researchers to design novel interfaces to support highlighting and comparing important features among instances with tabular data. However, such novel XAI interface is not applicable for image data, and new XAI algorithms will need to be proposed, such as in the work of Chen et al. [27] and Codella et al. [31].

- Participants suggested clicking features in feature attribution to check details of feature shape. It can be regarded as a combination of the two explanatory forms. Such a combination can be achieved at the interface level, e.g.: Gamut [44], or at the algorithmic level, e.g.: COGAM [12].

- The advantages and disadvantages of similar example and typical example seem to be complementary to each other. A new type of example-based explanation may be proposed accordingly: it creates typical examples that are representative to the target class, while being as similar as possible to the input instance. Thus, by taking the advantage of similar example, it is similar to the input instance to be easily understood, thus overcoming the disadvantage of typical example for being unrelated to the input instance; meanwhile it inherits the advantage of typical example for being distinctive and not confusing.

The above examples illustrate using the explanatory forms as building blocks to create novel XAI algorithms/interfaces. In addition to the identified design implications in our study, XAI researchers can use the above prototyping method to identify users' requirements on their particular tasks, and propose new interfaces or algorithmic solutions accordingly.

Since explanation is a social process [71], an advanced XAI system may be trained to construct an explanation dialog [91] for personalized explanation [83] that mimics the human explanation process. New XAI algorithms can also be created in line with such a manner, for example, by using reinforcement learning to use different explanatory forms to respond to user's current query, and make the explanation adapted to users' preferences or current explanation goals. We made the data collected from the user study publicly available (Supplementary Material S3: EUCA dataset) to facilitate XAI research in this direction.

### 7.2 End-Users' Explanation Goals

XAI techniques are abundant, but the understanding of end-users' needs is little. In this section, we discuss the user study findings on end-users' diverse explanation goals. Our findings reveal two major categories of end-users' need for explainability: explanation for the verification/justification of AI model decisions, and explanation for betterment.

*7.2.1 "How do I know when my prediction is not an error?": End-users' Verification of Decision Quality.* After acknowledging AI's decision is probabilistic, most users need explanations for decision verification, so that they could confidently incorporate AI into their own decision process on high-stakes tasks. In our study, we discovered users frequently seek decision quality metrics, followed by explanation answering *why* or *how* questions (such as feature attribution and similar example) to verify the decisions. Users usually request such information **1**) during the initial deployment stage [60] when trust has not been established. In this phase, users do not have prior experience with the AI (i.e.: lack knowledge on the observed performance [99]), as stated in the explanation goals trust, safety and communication; and **2**) when AI's decision is being challenged, as stated in the explanation goals detecting bias and unexpected prediction.

Our quantitative results also revealed that the decision quality-related metrics (output, performance, and dataset) were frequently selected and ranked higher for the above explanation goals.

In our study, we found that most participants accepted and understood the decision certainty in output, followed by performance. The training data distribution in a dataset was the least comprehensible form. Interpreting these metrics may require a certain degree of data analytic skills and could be time-consuming. And different numbers may lead to contradictory impressions of the model decision quality and may cause users' frustration. Thus, in real-world applications, it may not be feasible for end-users to check all the metrics. To indicate the probabilistic nature of AI, based on our findings on output (Section 5.10), we suggest a possible workaround: to provide the range of prediction on-demand or a point prediction within its range. The range may bring additional benefits of leaving rooms for flexible and negotiable decisions for specific tasks. Another suggestion is to provide a unified and precise uncertainty estimation [18] metric that is case specific, incorporating all sources of decision uncertainty (such as performance on model capabilities, prior knowledge about the training data distribution, noise on input data, etc), to indicate the capabilities and limitations of AI prediction.

Alongside the above metrics, various other explanatory forms were selected by participants to verify AI's decision (trust, safety, bias, unexpected). The selection of explanatory forms largely depends on the specific task, the explanation goal, and users' preferences, and there are no definite patterns. Previous quantitative results showed discrepancies of providing local explanations (feature attribution or similar example) and its effect on trust calibration and users' decision accuracy [57, 104], indicating explanations may play a complex role in the AI-assisted decision process. It may involve complex interactions among factors such as users' perception of the explanatory forms and their visual representations/layouts, AI and human's different error zones [104], explanatory information overload, users' cognitive bias when interpreting the explanations [61], and how faithful the explanations are to the underlying AI model [50], etc. Future research is needed to explore these factors and their effects on human-AI collaborative decision quality. Because there lacks a universal model to predict various explanatory forms and their outcomes, our proposed EUCA framework could serve as a practical prototyping tool to quickly test the effects of various explanatory forms to guide the design process.

*7.2.2 Explanations for Betterment.* The other major motivation to check AI's explanation is to move beyond decision verification, and to improve users' task performance either in the short run or long run, such as to improve the predicted outcome, enhance users' learning and problem-solving skills, discover new knowledge, and trade-off among multiple objectives. Those explanation goals may emerge as users established trust and adopted AI into their decision workflow. As AI surpasses human performance in some critical tasks, AI can act as a knowledgeable source providing insights for humans to improve our own welfare. Although research in this direction is relatively limited, some prior works provide promising results on using machine explanations to improve users' knowledge and task performance [14, 56].

*7.2.3 Optimizing and evaluating XAI for end-user-oriented explanation goals.* Existing XAI design and evaluation objectives are ill-defined with an engineer-centered perspective, such as for building trust, model understanding and debugging [28]. Our user study provides a thorough understanding of various explanation goals from end-user perspective. This specifies and extends existing XAI design objectives by maximizing the utility of XAI in real-world end-user-oriented applications. It inspires the research community to design and optimize XAI that targets end-users explanation goals from model decision verification, to improving human-XAI collaborative task performance. For example, based on findings from the EUCA user study and the study with clinical end-users of neurosurgeons, Jin

46 W. Jin, et al.et al. formulated the design and evaluation objective for clinical XAI according to the explanation goal of decision verification [49]. The corresponding XAI evaluation metric they proposed, informative plausibility, measures whether an explanation is indicative of model decision quality by end-users' judgment on explanation plausibility. The EUCA framework is a call for more research in the direction of end-user-oriented XAI design.

## 8   LIMITATIONS AND FUTURE WORK

We summarized the end-user-friendly explanatory forms from technically achievable solutions via a literature review process. We aim to include the majority of existing explanatory forms using the information saturation criterion: i.e., no more additional explanatory forms could be identified. This process manifested in a conceptual model of the 12 end-user-friendly explanatory forms that served as a starting point for the subsequent user study [39]. We did not aim to conduct an exhaustive, comprehensive systematic review, which is beyond what one paper could achieve. Since XAI techniques are fast evolving, the current framework may not necessarily cover all possible algorithms. The EUCA framework is a moderate initial step towards a practical end-user-centered XAI framework. It is upgradable with any emerging XAI technologies on the EUCA website[3].

Due to the high-stakes nature and the current relatively limited adoption of AI in critical decision-support, it is challenging to gain access to real-world AI systems in high-stake facilities (such as police offices, courts, clinics/hospitals, banks) to conduct user studies on multiple critical tasks, and recruit domain-specific end-users (such as physicians, police officers, judges, bankers). This is beyond the scope of what one single paper could achieve. Therefore, in the user study, we designed four fictional vignettes to represent the variability of AI-supported critical decision-making tasks, and participants' responses were based on conjecture rather than their real experience with AI.

In our ongoing future work, we are collaborating with physicians as domain expert end-users. To design and implement a novel XAI system on medical image-related tasks for clinical decision support, we conducted one-on-one interviews with 25 doctors and used the EUCA framework to co-design prototypes with them. Other future work may apply the EUCA framework in other domain-specific XAI design and development practices to iterate and improve the framework.

Bias may be involved in the card selection & sorting of explanatory forms, as we noticed a few participants selected a card because it contained certain features rather than its distinguished form, despite in the follow-up questions we asked "what if a specific feature was or wasn't included". The explanatory forms, their contents, the particular visual representations, tasks, user's current explanation goals, and user types all played a role in participants' selection of explanatory forms, and our study design could not disentangle them. The quantitative results from card selection & sorting are meant to serve as a reference only. They are not meant to be used directly to choose explanatory forms without the prototyping process, due to the above complex factors involved.

Future work may design randomized controlled user studies to quantitatively examine the effects of the above factors in detail to guide the choice of explanatory forms in specific contexts.

---

[3]http://weinajin.github.io/end-user-xai

## 9 CONCLUSION

Designing end-user-oriented explainable AI systems faces many challenges. From the user's perspective, **1**) End-users have diverse roles, tasks, and explanation goals. **2**) End-users lack technological knowledge, which is a prerequisite for some XAI systems in order to interpret the explanation. From the XAI practitioner's perspective, **3**) practitioners' expertise on AI or HCI/UIUX design usually does not overlap, and there lacks boundary objects to connect the two fields and facilitate collaboration between AI and HCI practitioners. **4**) There lacks tools to support UI/UX design, prototyping, and co-design process.

To address the above challenges, we developed the end-user-centered XAI framework EUCA with a collaborative effort of combining AI and HCI expertise. EUCA considers not only the human-centered perspective but also the technological capabilities, so that the design solutions are both end-user-oriented and technically achievable. It acts as a boundary object between AI and HCI fields and provides supports for UI/UX design, prototyping, and co-design process.

To apply EUCA in practice, XAI designers can use the provided design templates/examples to create prototyping cards for the twelve explanatory forms. The explanatory forms are end-user-friendly and are identified from a technically achievable solution space. They are a familiar and common language to both end-users and XAI practitioners. With the prototyping cards, designers can conduct a participatory design process to take stakeholders' input and iterate the prototype. The stakeholders can comment, sort, combine, and revise the prototyping cards, to use them as building blocks to build a low-fidelity prototype. Designers can also refer to the user study findings to be informed by end-users about the properties of the explanation forms (their strength, weakness, UI/UX design implications, and applicable explanation goals), and to understand end-users' diverse explanation goals. The corresponding XAI algorithms for each explanatory form can facilitate developers to implement a functional prototype.

As an initial step towards end-user-centered XAI, the EUCA framework provides a practical prototyping tool. It helps HCI/AI practitioners and researchers to understand users' needs for explainability and develop end-user-oriented XAI systems accordingly.

## ACKNOWLEDGMENTS

We thank all study participants for their time, effort, and valuable inputs in the study. We thank Sheelagh Carpendale, Parmit Chilana, Ben Cardoen, Pegah Kiaei, Zipeng Liu, and Mingbo Cai for the helpful discussions in shaping this work. We thank all reviewers for their valuable comments. The first author was supported by Simon Fraser University Big Data Initiative The Next Big Question Funding. The first author would like to appreciate her family for their generous support to complete the work during the difficult times in 2020.

EUCA: the End-User-Centered XAI Framework 51[79] Marco Tulio Ribeiro, Sameer Singh, and Carlos Guestrin. 2016. "Why Should I Trust You?": Explaining the Predictions of Any Classifier. In *Proceedings of the 22nd ACM SIGKDD International Conference on Knowledge Discovery and Data Mining* (San Francisco, California, USA) *(KDD '16)*. Association for Computing Machinery, New York, NY, USA, 1135–1144. https://doi.org/10.1145/2939672.2939778

[80] Marco Tulio Ribeiro, Sameer Singh, and Carlos Guestrin. 2018. Anchors: High-Precision Model-Agnostic Explanations. In *AAAI Conference on Artificial Intelligence (AAAI)*.

[81] Mireia Ribera and Agata Lapedriza. 2019. Can we do better explanations? A proposal of user-centered explainable AI. In *Joint Proceedings of the ACM IUI 2019 Workshops*. http://ceur-ws.org/Vol-2327/IUI19WS-ExSS2019-12.pdf

[82] Mary Beth Rosson and John M. Carroll. 2002. *Scenario-Based Design*. L. Erlbaum Associates Inc., USA, 1032–1050.

[83] Johanes Schneider and Joshua Peter Handali. 2019. Personalized explanation in machine learning. *ArXiv* abs/1901.00770 (2019).

[84] Karen Simonyan, Andrea Vedaldi, and Andrew Zisserman. 2013. *Deep Inside Convolutional Networks: Visualising Image Classification Models and Saliency Maps*. Technical Report. arXiv:1312.6034v2 https://arxiv.org/abs/1312.6034v2

[85] Alison Smith-Renner, Ron Fan, Melissa Birchfield, Tongshuang Wu, Jordan Boyd-Graber, Daniel S. Weld, and Leah Findlater. 2020. No Explainability without Accountability. In *Proceedings of the 2020 CHI Conference on Human Factors in Computing Systems*. Association for Computing Machinery (ACM), New York, NY, USA, 1–13. https://doi.org/10.1145/3313831.3376624

[86] Kacper Sokol and Peter Flach. 2020. Explainability fact sheets: A framework for systematic assessment of explainable approaches. *FAT* 2020 - Proceedings of the 2020 Conference on Fairness, Accountability, and Transparency* (2020), 56–67. https://doi.org/10.1145/3351095.3372870 arXiv:1912.05100

[87] Sarah Tan, Rich Caruana, Giles Hooker, Paul Koch, and Albert Gordo. 2018. *Learning Global Additive Explanations for Neural Nets Using Model Distillation*. Technical Report. arXiv:1801.08640v2 https://youtu.be/ErQYwNqzEdc.

[88] Amy Turner, Meena Kaushik, Mu-Ti Huang, and Srikar Varanasi. 2020. Calibrating Trust in AI-Assisted Decision Making. (2020).

[89] Danding Wang, Qian Yang, Ashraf Abdul, and Brian Y. Lim. 2019. Designing Theory-Driven User-Centric Explainable AI. In *Proceedings of the 2019 CHI Conference on Human Factors in Computing Systems* (Glasgow, Scotland Uk) *(CHI '19)*. Association for Computing Machinery, New York, NY, USA, 1–15. https://doi.org/10.1145/3290605.3300831

[90] Jonas Wanner and Christian Janiesch. 2020. How much is the black box? The value of explainability in machine learning models. (2020).

[91] Daniel S. Weld and Gagan Bansal. 2019. The challenge of crafting intelligible intelligence. *Commun. ACM* 62, 6 (mar 2019), 70–79. https://doi.org/10.1145/3282486 arXiv:1803.04263

[92] P. Welinder, S. Branson, T. Mita, C. Wah, F. Schroff, S. Belongie, and P. Perona. 2010. *Caltech-UCSD Birds 200*. Technical Report CNS-TR-2010-001. California Institute of Technology.

[93] Christine T. Wolf. 2019. Explainability scenarios: Towards scenario-based XAI design. *International Conference on Intelligent User Interfaces, Proceedings IUI* Part F1476 (2019), 252–257. https://doi.org/10.1145/3301275.3302317

[94] Jennifer Wortman Vaughan and Hanna Wallach. [n.d.]. A Human-Centered Agenda for Intelligible Machine Learning. *Jennwv.Com* ([n. d.]). http://www.jennwv.com/papers/intel-chapter.pdf

[95] Hongyu Yang, Cynthia Rudin, and Margo Seltzer. 2017. Scalable Bayesian Rule Lists. In *Proceedings of the 34th International Conference on Machine Learning - Volume 70* (Sydney, NSW, Australia) *(ICML'17)*. JMLR.org, 3921–3930.

[96] Qian Yang. 2018. Machine Learning as a UX Design Material: How Can We Imagine Beyond Automation, Recommenders, and Reminders? https://aaai.org/ocs/index.php/SSS/SSS18/paper/view/17471

[97] Qian Yang, Alex Scuito, John Zimmerman, Jodi Forlizzi, and Aaron Steinfeld. 2018. Investigating How Experienced UX Designers Effectively Work with Machine Learning. In *Proceedings of the 2018 Designing Interactive Systems Conference* (Hong Kong, China) *(DIS '18)*. Association for Computing Machinery, New York, NY, USA, 585–596. https://doi.org/10.1145/3196709.3196730

[98] Qian Yang, Aaron Steinfeld, Carolyn Rosé, and John Zimmerman. 2020. Re-Examining Whether, Why, and How Human-AI Interaction Is Uniquely Difficult to Design. In *Proceedings of the 2020 CHI Conference on Human Factors in Computing Systems* (Honolulu, HI, USA) *(CHI '20)*. Association for Computing Machinery, New York, NY, USA, 1–13. https://doi.org/10.1145/3313831.3376301

[99] Ming Yin, Jennifer Wortman Vaughan, and Hanna Wallach. 2019. Understanding the Effect of Accuracy on Trust in Machine Learning Models. In *Proceedings of the 2019 CHI Conference on Human Factors in Computing Systems* (Glasgow, Scotland Uk) *(CHI '19)*. Association for Computing Machinery, New York, NY, USA, 1–12. https://doi.org/10.1145/3290605.3300509

[100] Fisher Yu, Wenqi Xian, Yingying Chen, Fangchen Liu, Mike Liao, Vashisht Madhavan, and Trevor Darrell. 2018. BDD100K: A Diverse Driving Video Database with Scalable Annotation Tooling. *CoRR* abs/1805.04687 (2018). arXiv:1805.04687 http://arxiv.org/abs/1805.04687

[101] Kun Yu, Shlomo Berkovsky, Ronnie Taib, Jianlong Zhou, and Fang Chen. 2019. Do I Trust My Machine Teammate? An Investigation from Perception to Decision. In *Proceedings of the 24th International Conference on Intelligent User Interfaces* (Marina del Ray, California) *(IUI '19)*. Association for Computing Machinery, New York, NY, USA, 460–468. https://doi.org/10.1145/3301275.3302277

[102] Quanshi Zhang, Yu Yang, Haotian Ma, and Ying Nian Wu. 2018. Interpreting CNNs via Decision Trees. (jan 2018). arXiv:1802.00121 http://arxiv.org/abs/1802.00121

[103] Yu Zhang, Peter Tiňo, Aleš Leonardis, and Ke Tang. 2021. A Survey on Neural Network Interpretability. arXiv:2012.14261 [cs.LG]

[104] Yunfeng Zhang, Q. Vera Liao, and Rachel K.E. Bellamy. 2020. Efect of confidence and explanation on accuracy and trust calibration in AI-assisted decision making. *FAT* 2020 - Proceedings of the 2020 Conference on Fairness, Accountability, and Transparency* (2020), 295–305. https://doi.org/10.1145/3351095.3372852 arXiv:2001.02114